\documentclass[floatfix,twocolumn,showpacs,preprintnumbers,amsmath,amssymb,pra,superscriptaddress,longbibliography]{revtex4-1}
\usepackage{color}
\usepackage[usenames,dvipsnames,svgnames,table]{xcolor}
\usepackage[colorlinks=true,linkcolor=blue,urlcolor=blue,citecolor=blue]{hyperref}

\usepackage{mathtools}
\usepackage{graphicx}
\usepackage{dcolumn}
\usepackage{array}
\usepackage{lipsum}
\usepackage{bm}
\usepackage{subfigure}
\usepackage{amssymb}
\usepackage{multirow}
\usepackage{tabularx}
\usepackage{amsmath}
\usepackage{braket}

\renewcommand{\vec}[1]{\mathbf{#1}}

\begin{document}

\title{Averaging over atom snapshots  in linear-response TDDFT of disordered systems:\\
A case study of warm dense hydrogen}

\author{Zhandos A. Moldabekov}
\email{z.moldabekov@hzdr.de}
\affiliation{Center for Advanced Systems Understanding (CASUS), D-02826 G\"orlitz, Germany}
\affiliation{Helmholtz-Zentrum Dresden-Rossendorf (HZDR), D-01328 Dresden, Germany}




\author{Jan Vorberger}
\affiliation{Helmholtz-Zentrum Dresden-Rossendorf (HZDR), D-01328 Dresden, Germany}

\author{Mani Lokamani}
\affiliation{Information Services and Computing, Helmholtz-Zentrum Dresden-Rossendorf (HZDR), D-01328 Dresden, Germany}

\author{Tobias Dornheim}
\affiliation{Center for Advanced Systems Understanding (CASUS), D-02826 G\"orlitz, Germany}
\affiliation{Helmholtz-Zentrum Dresden-Rossendorf (HZDR), D-01328 Dresden, Germany}

\begin{abstract}
Linear-response time-dependent density functional theory (LR-TDDFT) simulations of disordered extended systems require averaging over different snapshots of ion  configurations to minimize finite size effects due to the snapshot--dependence of the electronic density response function and related  properties. We present a consistent scheme for the computation of the macroscopic Kohn-Sham (KS) density response function connecting an average over snapshot values  of charge density perturbations to the averaged values of  KS potential variations. This allows us to formulate the LR-TDDFT within the adiabatic (static) approximation for the exchange-correlation (XC) kernel  for disordered systems, where the static XC kernel is computed using the direct perturbation method [Moldabekov {\textit et al.} J. Chem. Theory Comput. {\bf 19}, 1286 (2023)].  The presented approach allows one to compute the macroscopic dynamic density response function as well as the dielectric function with a static XC kernel generated for  any available XC functional. The application of the developed workflow is demonstrated for the example of warm dense hydrogen. The   presented approach is applicable for various types of extended disordered systems such as warm dense matter, liquid metals, and dense plasmas.   

\end{abstract}

\maketitle

\section{Introduction} 

\textit{Ab intio} simulations play an indispensable role in the understanding of the physics and chemistry of materials at extreme conditions. Such warm dense matter (WDM) naturally appears in the interiors of planets \cite{saumon1}, brown dwarfs \cite{becker}, and white dwarfs \cite{CBFS00}, and in the outer layer of relatively cold neutron stars \cite{Daligault_2009}. In experiments, WDM is created using powerful lasers and shock compression at facilities such as the National Ignition Facility (NIF) \cite{Hayes2020, Moses_NIF, Chapman_POP_2014} and the European X-ray Free-Electron Laser (XFEL) \cite{Zastrau2021, Zastrau}. Without accurate simulations, the extreme conditions and the short life time of WDM states generated in experiments often hinder effective diagnostics. Indeed, one usually has to rely on simulations to extract properties (structural factors, free energy etc.) from the experimental data.

One of the common diagnostic tools in WDM experiments is X-ray Thomson scattering (XRTS) \cite{Glenzer_RevModPhys}, which provides information about the dynamical structure factor  of the electrons $S(\vec q, \omega)$. To describe the XRTS signal and, in this way, extract the maximum amount of information about the dynamic properties of WDM, one needs accurate simulation results for $S(\vec q, \omega)$. By virtue of the fluctuation--dissipation theorem connecting $S(\vec q, \omega)$ with the dynamical linear density response function \cite{quantum_theory},  agreement between experiment and theory for the XRTS signal 
can provide reliable access to a great variety of dynamical properties such as the dynamical dielectric function, conductivity, and energy loss characteristics.  
Furthermore, accurate simulations can be used to guide and design future experiments. This is particularly important for highly challenging tasks like the development of inertial confinement fusion (ICF) technology. 

Commonly used \textit{ab initio} methods for the computation of the dynamical structure factor are linear-response time-dependent density functional theory (LR-TDDFT) and real-time time-dependent density functional theory (RT-TDDFT) \cite{POP_review}, which is formally equivalent in the linear-response regime to LR-TDDFT \cite{book_Ullrich}. More recently, the imaginary time density--density correlation function [a two-sided Laplace transform of $S(\vec q, \omega)$] that can, in principle, be computed from highly accurate quantum Monte Carlo (QMC) methods has been brought forward by Dornheim \textit{et al.} \cite{Dornheim_insight_2022, Dornheim_moments_2022, Dornheim_T_2022, Dornheim_PTR_2022} as a new tool for the investigation of the dynamical properties of WDM in thermodynamic equilibrium and beyond~\cite{Vorberger_PRX_2023}. All of these methods have certain computational bottlenecks with respect to the number of particles within the simulation. This can lead to finite size effects that have to be minimized  to increase the accuracy of the results. 
For example, a too small number of particles clearly leads to errors
in the calculations of thermodynamic properties (pressure, energy \textit{etc}) \cite{Brown_PRL_2013,Holzmann_PRB_2016,dornheim_prl,review,Dornheim_JCP_2021} as well as dynamic properties such as dynamic density response function \cite{Dornheim_PRE_2020, Moldabekov_SciRep_2022}. 

In this work, we consider finite size effects in the electronic density response function of disordered systems due to the dependence on the positions of the nuclei in a snapshot. 
This problem vanishes for crystals if the box length is commensurate with the crystal periodicity. In this case, periodic boundary conditions represent a real physical picture of solids . In the high temperature limit, which corresponds to the fully ionized plasma state, the electrons are free and the ions provide a neutralizing background \cite{Chabrier_1990, zhandos1}. 
The finite size effect due to the dependence on the used ionic snapshot is expected to be strong for extended disordered systems with sufficiently strong electron-ion coupling. This is often the case for WDM.  

For disordered systems, a standard way to reduce finite size effects is to perform an averaging of the simulation results over different snapshots. Alternatively, one can evaluate the uncertainty due to the snapshot--dependence  by investigating the properties of interest at different values of the number of particles. In this work, we analyse the effectiveness of these strategies for WDM by considering the density response function of warm dense hydrogen. We consider the dynamic and static density response function, the Kohn-Sham (KS) response function,  and the static exchange-correlation kernel; the latter is the second order variational derivative of the XC functional with respect to the density~\cite{marques2012fundamentals}. 

Recently, Moldabekov \textit{et al} \cite{Moldabekov_dft_kernel, moldabekov_lr-tddft_2023} have presented an approach that allows one to compute the static XC kernel for any available XC functional on any rung of Jacob's ladder~\cite{Perdew_AIP_2001} without explicitly performing  the cumbersome second order functional derivative. The key idea of the method is to compute the density change due to the external static harmonic perturbation. This method was used to quantify the quality of various XC functionals  by comparing with exact  QMC data for warm dense hydrogen \cite{Moldabekov_dft_kernel, moldabekov_lr-tddft_2023} and the uniform electron gas \cite{Moldabekov_dft_kernel, Moldabekov_non_empirical_hybrid, hybrid_results, moldabekov2021relevance, Moldabekov_PRB_2022}. One of the remaining open questions regarding the application of the direct perturbation method for the computation of the static XC kernel had been the excitation of density perturbations at wavenumbers different from the wavenumber of the external harmonic perturbation, which vanishes only in the thermodynamic limit $N\to \infty$ (with $N$ being the number of particles).  This is one of the problems considered in this work.  

The static XC kernel computed using the direct perturbation approach requires knowledge of the macroscopic static KS response function \cite{moldabekov_lr-tddft_2023}. Furthermore, for the application of this static XC kernel in LR-TDDFT, one needs information about the macroscopic dynamic KS response function.
Therefore, an open question is how to average the KS response function over different snapshots. The naive way is to perform arithmetic averaging over the KS response functions of individual snapshots.
Here we show that this does not follow from the formal definition of the KS response function.
This is a consequence of the nonlinear dependence of the linear density response on the KS response function.
As a suitable alternative, we present a rigorously derived formula for the proper averaging the  macroscopic dynamic KS response function  over snapshots. 

The presented results are relevant not only for the density functional theory (DFT) of WDM, but also for the other simulation methods of WDM, such as QMC \cite{Bohme_PRL_2022, Bohme_PRE_2022}. 
Furthermore, the presented analysis of the finite size effects are relevant for the simulations of other disordered  systems like liquid metals.

In the next section \ref{s:theory}, we present the theory for computing the averaged density response properties and  the corresponding formulation of an LR-TDDFT based calculation scheme of the macroscopic density response function.
The simulation details are provided in Sec. \ref{s:details}. The application of the developed computational scheme is demonstrated in Sec. \ref{s:results} for the example of warm dense hydrogen.  
We conclude the paper by summarizing  the results and providing an outlook in Sec. \ref{s:end}.


\section{Theory}\label{s:theory}

We consider LR-TDDFT with an adiabatic exchange--correlation kernel. 
One of the commonly used approximations for a static XC kernel  is the adiabatic local density approximation (ALDA). 
For extended systems, the approach based on the explicit implementation of the second order functional derivative of an XC functional with respect to the density is currently restricted to the ALDA and adiabatic generalized gradient approximation (AGGA). In contrast, the direct perturbation approach is capable of computing the static (adiabatic) XC kernel for any available XC functional from LDA all the way across Jacob's Ladder to hybrid XC functionals \cite{Moldabekov_dft_kernel, hybrid_results, Moldabekov_non_empirical_hybrid}.


In this section, we first consider the direct perturbation approach to compute the static XC kernel. For this purpose, the static density response function and the static KS response function must be computed by comparing the perturbed and unperturbed density and KS potential values, respectively. Therefore, we discuss how the averaging over snapshots is performed for these quantities. After that, we discuss the LR-TDDFT approach to disordered systems with an adiabatic exchange--correlation kernel. We provide a scheme for the computation of the dynamic macroscopic KS response function in LR-TDDFT that represents a properly averaged value over snapshots.  We show that it is not equivalent to the arithmetic mean of the KS response functions computed for separate snapshots. Using  a consistent scheme for the averaging, we discuss how a dynamic macroscopic KS response function can be combined with the static XC kernel from the direct perturbation approach to compute the dynamic macroscopic density response function.


\subsection{The direct perturbation approach}\label{s:theory_A}
\subsubsection{Static total density response to a bare external perturbing field}
To obtain the static density response function $\chi(\mathbf{q})=\chi(\mathbf{q}, \omega=0)$, we perform two sets of KS-DFT simulations.
First, we find equilibrium density values of electrons in the field of the ions (the unperturbed system)  and then we repeat the simulation, applying an extra static harmonic field (the perturbed system). The corresponding  Hamiltonian reads
\begin{eqnarray}\label{eq:Hamiltonian_modified}
 \hat{H}_{\mathbf{q},A} = \hat{H}_e + 2 A \sum_{j=1}^N \textnormal{cos}\left( \mathbf{q}\cdot\hat{\mathbf{r}}_j \right)\ ,
\end{eqnarray}
where  $\hat{H}_e$ is the Hamiltonian of the unperturbed system, and $A$ and $\vec q$ being the amplitude and the wavevector of the external perturbation. 

In the case of the unperturbed system ($A=0$), the density distribution of the electrons ${n_{A=0}^{i}(\mathbf{r})}$ for a given spatial configuration of ions  is not uniform (with $i$ being the label of a particular snapshot), i.e. ${n_{A=0}^{i}(\mathbf{r})}\neq n_0 = {const}$, where $n_0$ is the mean value of the density.
In contrast to crystals, the density distributions for different ionic snapshots are not equivalent for disordered systems.
Indeed, the averaged value of the density distribution tends to a constant  for a large number of snapshots $N_s$
\begin{equation}\label{eq:n0}
    \braket{n_e(\mathbf{r})}_{A=0}=\lim_{N_s \to \infty}\frac{1}{N_s}\sum_{i=1}^{N_s} {n_{A=0}^{i}(\mathbf{r})} = n_0.
\end{equation}

Physically, the homogeneity of disordered systems means that diagnostics (e.g. XRTS) is performed on a macroscopic sample, which has properties independent of the probing direction.

The application of the external harmonic field according to Eq. (\ref{eq:Hamiltonian_modified}) leads to the density perturbation
\begin{equation}
\Delta n_{\mathbf{q},A}^{i}(\mathbf{r})={n_{A}^{i}(\mathbf{r})}-{n_{A=0}^{i}(\mathbf{r})}.    
\end{equation}

 If the perturbation amplitude $A$ is small enough, the non-linear response can be neglected \cite{Dornheim_PRR_2021, Moldabekov_JCTC_2022, Dornheim_PRL_2020} and  $\Delta n_{\mathbf{q},A}^{i}(\mathbf{r})$ can be described by linear response theory (LRT). Due to periodic boundary conditions, the density perturbation can be written as a Fourier series.
Since a cosine perturbation is applied in Eq. (\ref{eq:Hamiltonian_modified}), here we use a Fourier cosine series,
 \begin{eqnarray}\label{eq:expansion}
 \Delta n_{\mathbf{q},A}^{i}(\mathbf{r}) = 2 \sum_{\vec G}
{\rho_{\vec G}^{i}(\vec q)}
 \textnormal{cos}\big(
 \left(\mathbf{q}+\mathbf{G}\right)\cdot\mathbf{r}
 \big)\ ,
 \end{eqnarray}
where $\vec G$ is the reciprocal lattice vector and the factor two is conventional (cf. the perturbation term in Eq. (\ref{eq:Hamiltonian_modified})). 


In the case of a harmonic perturbation of a uniform system---such as the uniform electron gas---the density perturbation has the same wavenumber as the external harmonic perturbation, i.e., only $  {\rho_{\vec G=0}^{i}(\vec q)}$ is non-zero \cite{dornheim_pre, Dornheim_PRR_2021}. 
Similarly, after averaging over snapshots, only the  term with $\vec G=0$ should survive for disordered systems,
\begin{equation}\label{eq:delta_n}
\begin{split}
     \Delta n(\mathbf{r})_{\mathbf{q}, A} &= \lim_{N_s \to \infty} \frac{1}{N_s}\sum_{i=1}^{N_s} \Delta n_{\mathbf{q},A}^{i}(\mathbf{r})\\
     &=
     2\left(\lim_{N_s \to \infty} \frac{1}{N_s} \sum_{i=1}^{N_s} {\rho_{\vec G=0}^{i}(\vec q)}\right)  \textnormal{cos}\left(
\mathbf{q}\cdot\mathbf{r}
 \right)\\
 &=2\braket{\rho(\vec q)}_{\vec G=0}  \textnormal{cos}\left(
\mathbf{q}\cdot\mathbf{r}
 \right)\ ,
\end{split}
\end{equation}

and all terms with $\vec G\neq 0$ vanish after averaging,

\begin{equation}\label{eq:rho_to_0}
    \braket{\rho(\vec q)}_{\vec G\neq0}=\lim_{N_s \to \infty} \frac{1}{N_s}\sum_{i=1}^{N_s} {\rho_{\vec G\neq0}^{i}(\vec q)} =0\ .
\end{equation}

The validity of Eq. (\ref{eq:n0}), Eq. (\ref{eq:delta_n}), and Eq. (\ref{eq:rho_to_0})  is demonstrated numerically in Sec. \ref{s:results} for the example of warm dense hydrogen by computing the averaged values of the unperturbed and perturbed density.

The density response function relates the density perturbation to the external harmonic perturbation in a linear fashion,
\begin{eqnarray}\label{eq:delta_n_LRT_G}
   \rho_{\vec G}^{i}(\vec q)  = 2 A \textnormal{cos}\left(\mathbf{q}\cdot\mathbf{r}\right)\chi_{\vec G}^{i}(\mathbf{q})\  .
\end{eqnarray}

Using $\Delta n_e(\mathbf{r})_{\mathbf{q}, A}$ from Eq. (\ref{eq:delta_n}) and Eq. (\ref{eq:delta_n_LRT_G}), we can write
\begin{equation}
    \Delta n(\mathbf{r})_{\mathbf{q}, A}=2A\textnormal{cos}\left(\mathbf{q}\cdot\mathbf{r}
 \right)\frac{1}{N_s}\sum_{i=1}^{N_s} \chi_{\vec G=0}^{i}(\mathbf{q})\ .
\end{equation}
Therefore, one can compute the macroscopic static linear density response function $\chi(\mathbf{q})$ of a disordered system (that is homogeneous on average) according to the relation
\begin{eqnarray}\label{eq:delta_n_LRT}
   \Delta n(\mathbf{r})_{\mathbf{q},A}  = 2 A \textnormal{cos}\left(\mathbf{q}\cdot\mathbf{r}\right)\chi(\mathbf{q})\ ,
\end{eqnarray}
where 
\begin{equation}\label{eq:my_chi}
    \chi(\mathbf{q})=\frac{1}{N_s}\sum_{i=1}^{N_s} \chi_{\vec G=0}^{i}(\mathbf{q})\ .
\end{equation}

As follows from Eq. (\ref{eq:my_chi}), one can compute  $\chi(\mathbf{q})$ using the density perturbation values computed for individual snapshots,
\begin{equation}\label{eq:sum_chi}
\begin{split}
     \chi(\mathbf{q}) &=\frac{1}{N_s}\sum_{i=1}^{N_s} \frac{{\rho_{\vec G=0}^{i}(\vec q)}}{A} =\frac{\braket{\rho(\vec q)}_{\vec G=0}}{A}\ .
\end{split}
\end{equation}

From Eq. (\ref{eq:rho_to_0}), it follows that, on average, one has

\begin{equation}\label{eq:chi_to_0}
    \lim_{N_s \to \infty} \frac{1}{N_s}\sum_{i=1}^{N_s} \frac{{\rho_{\vec G\neq 0}^{i}(\vec q)}}{A}=\lim_{N_s \to \infty} \frac{1}{N_s}\sum_{i=1}^{N_s} {\chi_{\vec G\neq0}^{i}(\vec q)} =0\ .
\end{equation}

\subsubsection{Static KS response function from the direct perturbation method}

In the case of the unperturbed system with $A=0$, the KS potential for a given snapshot of ionic positions is inhomogeneous, i.e.
${v_{\rm KS,~A=0}^{i}(\mathbf{r})}\neq {const}$. Similarly to the electron density, the inhomogeneity  in the KS potentials of different snapshots vanishes upon averaging over snapshots. In the limit of an infinite number of snapshots $N_s$, the mean value of the KS potential becomes a constant,    

\begin{equation}\label{eq:v_ks_0}
    \braket{v_{\rm KS}(\mathbf{r})}_{A=0}=\lim_{N_s \to \infty}\frac{1}{N_s}\sum_{i=1}^{N_s} {v_{\rm KS,~A=0}^{i}(\mathbf{r})} = v_{\rm KS}^{0}\ .
\end{equation}

For a given snapshot $i$, the static cosine perturbation in Eq. (\ref{eq:Hamiltonian_modified}) leads to the perturbation of the KS potential

\begin{equation}
\Delta v_{\rm KS, A}^{i}(\mathbf{r})={v_{\rm KS, A}^{i}(\mathbf{r})}-{v_{\rm KS, A=0}^{i}(\mathbf{r})}\ ,    
\end{equation}
which we represent using a Fourier cosine series,
 \begin{eqnarray}\label{eq:expansion_vks}
 \Delta v_{\rm KS, A}^{i}(\mathbf{r}) = 2 \sum_{\vec G}
{u_{\vec G}^{i}(\vec q)}
 \textnormal{cos}\big(
 \left(\mathbf{q}+\mathbf{G}\right)\cdot\mathbf{r}
 \big)\ .
 \end{eqnarray}

Based on the same reasoning as for the electron density, we can write for the averaged value of the perturbation of the KS potential
\begin{equation}
\begin{split}
      \Delta v_{\rm KS}(\mathbf{r})_{\mathbf{q},A} &= \lim_{N_s \to \infty} \frac{1}{N_s}\sum_{i=1}^{N_s} \Delta v_{\rm KS, A}^{i}(\mathbf{r})\\
      &=2 \left(\lim_{N_s \to \infty} \frac{1}{N_s}\sum_{i=1}^{N_s} {u_{\vec G=0}^{i}(\vec q)}\right) \textnormal{cos}\left(
\mathbf{q}\cdot\mathbf{r}
 \right)\\
 &= 2 \braket{u(\vec q)}_{\vec G=0} \textnormal{cos}\left(
\mathbf{q}\cdot\mathbf{r}
 \right)\ ,
\end{split} \label{eq:delta_vks}
\end{equation}
where $ {u_{\vec G=0}^{i}(\vec q)}$ is the Fourier component of the KS potential perturbation at $\vec G=0$.

The static KS response function defines the response of the electron density to a change of the KS potential \cite{Kollmar}.
Using averaged values $\Delta n_{\mathbf{q},A}(\mathbf{r})$ from Eq. (\ref{eq:delta_n}) and  $\Delta v_{\rm KS}(\mathbf{r})_{\mathbf{q},A}$ from Eq. (\ref{eq:delta_vks}), the static KS response function connecting the  average electron density change and the average KS potential perturbation follows from the relation
\begin{eqnarray}\label{eq:delta_KS}
   \Delta n_{\mathbf{q},A}(\mathbf{r}) = \chi_{\rm KS}(\mathbf{q}) \Delta v_{\rm KS}(\mathbf{r})_{\mathbf{q},A}\ .
\end{eqnarray}

From Eq. (\ref{eq:delta_KS}) we find
\begin{equation}\label{eq:chi_KS}
    \chi_{\rm KS}(\mathbf{q})= \frac{\braket{\rho(\vec q)}_{\vec G=0}}{\braket{u(\vec q)}_{\vec G=0}}=\frac{\sum_{i=1}^{N_s}{\rho_{\vec G=0}^{i}(\vec q)}}{\sum_{i=1}^{N_s}{u_{\vec G=0}^{i}(\vec q)}}\ .
\end{equation}

Eq. (\ref{eq:chi_KS}) provides the macroscopic static KS response function connecting the average value of the electron density perturbation  to the average value of the perturbation of the KS potential over atomic snapshots. 

One of the conclusions following from Eq. (\ref{eq:chi_KS}) is that a direct average over KS response functions computed for individual snapshots is not a consistent way to deal with finite size effects. 
Indeed, one can formally compute the static KS response function for an individual snapshot as
\begin{equation}\label{eq:chi_ks_individual}
    \chi_{\rm KS, \vec G}^{i}(\mathbf{q}) =\frac{{\rho_{\vec G}^{i}(\vec q)}}{{u_{\vec G}^{i}(\vec q)}}\ ,
\end{equation}
and define the average value of the macroscopic KS response function as
\begin{equation}\label{eq:i_chi_KS}
    \braket{\chi_{\rm KS}(\mathbf{q})}= \frac{1}{N_s}\sum_{i=1}^{N_s} \chi_{\rm KS, \vec G=0}^{i}(\mathbf{q})\ .
\end{equation}
We observe that, in general, if ${n_{A=0}^{i}(\mathbf{r})}\neq n_0$ and ${v_{\rm KS,~A=0}^{i}(\mathbf{r})} \neq v_{\rm KS}^{0}$ for any snapshot, then $\chi_{\rm KS, \vec G}^{i}(\mathbf{q})$ defined by Eq. (\ref{eq:chi_KS})  and $\braket{\chi_{\rm KS}(\mathbf{q})}$ from Eq. (\ref{eq:i_chi_KS}) are not equivalent,

\begin{equation}\label{eq:neq}
\begin{split}
    \frac{\sum_{i=1}^{N_s}{\rho_{\vec G=0}^{i}(\vec q)}}{\sum_{i=1}^{N_s}{u_{\vec G=0}^{i}(\vec q)}}& \neq \frac{1}{N_s} \sum_{i=1}^{N_s} \frac{{\rho_{\vec G=0}^{i}(\vec q)}}{{u_{\vec G=0}^{i}(\vec q)}} \\
    &\Downarrow \\
    \chi_{\rm KS}(\mathbf{q})&\neq \braket{\chi_{\rm KS}(\mathbf{q})}\ .
\end{split}
\end{equation}

In a strict mathematical sense, the inequality (\ref{eq:neq}) is valid for any system that is not a perfect crystal (defect-free crystal).
This includes disordered systems such as warm dense matter and fluids.
Moreover, the inequality (\ref{eq:neq}) applies for solids with a large enough number of defects so that different 
snapshots are not equivalent. 

The reason to use Eq. (\ref{eq:chi_KS}) is because it is consistent with the linear response theory formulation for homogeneous systems. In contrast,  Eq. (\ref{eq:i_chi_KS}) results in inconsistency with the standard linear response theory for homogeneous systems. This is demonstrated in Appendix A.

\subsubsection{Static XC kernel}

The static XC kernel based on the averaged values over snapshots of the density and KS potential perturbations can be computed using the density response function $\chi(\mathbf{q})$ from Eq. (\ref{eq:sum_chi}) and the KS response function $\chi_{\rm KS}(\mathbf{q})$ from Eq. (\ref{eq:chi_KS})  \cite{Moldabekov_dft_kernel, moldabekov_lr-tddft_2023}:
\begin{equation}\label{eq:invert}
\begin{split}
     K_\textnormal{xc}(\mathbf{q}) &= -\left\{
 v(q) + \left( \frac{1}{\chi(\mathbf{q})} - \frac{1}{\chi_{\rm KS}(\mathbf{q})} \right)
 \right\}\ .
\end{split}
\end{equation}

In general, it follows from Eq. (\ref{eq:invert}) that
\begin{equation}
    K_\textnormal{xc}(\mathbf{q})\neq \frac{1}{N_s} \sum_{i=1}^{N_s} K_{\rm xc}^{i}(\vec q)\ ,
\end{equation}
where $K_{\rm xc}^{i}(\vec q)$ is computed for an individual snapshot:
\begin{equation}\label{eq:invert_i}
\begin{split}
     K_\textnormal{xc}^{i}(\mathbf{q}) &= -\left\{
 v(q) + \left( \frac{1}{\chi_{\vec G=0}^{i}(\mathbf{q})} - \frac{1}{\chi_{{\rm KS}, \vec G=0}^{i}(\mathbf{q})} \right)
 \right\}\ .
\end{split}
\end{equation}

Therefore, a straightforward arithmetic averaging using the static XC kernel for individual snapshots does not provide a consistent result.

The direct perturbation approach and Eq. (\ref{eq:invert}) allow one to compute the static XC kernel for any available XC functional.  It was used to compute the static XC kernel of the uniform electron gas and warm dense hydrogen (without averaging over snapshots) using ground state LDA, GGA, and meta-GGA functionals \cite{Moldabekov_dft_kernel, moldabekov_lr-tddft_2023} as well as a finite temperature LDA~\cite{groth_prl} in Ref.~\cite{Moldabekov_dft_kernel}. Furthermore, various hybrid XC functionals have been analyzed for the uniform electron gas both in the ground state and at high temperatures in Refs.~\cite{Moldabekov_non_empirical_hybrid, hybrid_results}.

\subsection{LR-TDDFT for disordered systems}\label{s:theory_B}

\subsubsection{Dynamic KS response function}

We note that $\chi_{\rm KS, \vec G=0}^{i}(\mathbf{q})$ and, in general, $\chi_{\rm KS, \vec G=0}^{i}(\mathbf{q},\omega)$ are well defined for both disordered systems and crystals and represent the macroscopic KS response function following from the macroscopic dielectric function defined within LR-TDDFT \cite{book_Ullrich, moldabekov_lr-tddft_2023}. In fact, the static macroscopic KS response function from LR-TDDFT and from the direct perturbation approach are equivalent to each other if the same XC functionals for both methods are used \cite{moldabekov_lr-tddft_2023}. To extend this equivalence to the dynamic case $\chi_{\rm KS, \vec G=0}^{i}(\mathbf{q},\omega)$, one needs to use a time-dependent perturbation generating correspondingly time-depended density  and KS potential perturbations in RT-TDDFT, which is outside of the scope of the present work. Nevertheless, LR-TDDFT and RT-TDDFT are formally equivalent in the linear response regime. Therefore, the equivalence of LR-TDDFT and the direct perturbation method is expected to hold for the dynamic case as well if the same XC functional and other parameters are used.

According to the definition of the macroscopic KS response function, we have for disordered systems,

\begin{equation}\label{eq:dyn_chi_KS}
    \braket{\rho(\vec q, \omega)}_{\vec G=0} = \chi_{\rm KS}(\mathbf{q}, \omega) \braket{u(\vec q, \omega)}_{\vec G=0}\ ,
\end{equation}
where 
\begin{equation}
    \braket{u(\vec q, \omega)}_{\vec G=0}=\frac{1}{N_s}\sum_i^{N_s}u_{\vec G=0}^{i}(\vec q, \omega)\ ,    
\end{equation}
and
\begin{equation}
    \braket{\rho(\vec q, \omega)}_{\vec G=0}=\frac{1}{N_s}\sum_i^{N_s} \rho_{\vec G=0}^{i}(\vec q, \omega)\ .
\end{equation}

Therefore,  the computation of  $\chi_{\rm KS}(\mathbf{q}, \omega)$ requires information about the dynamic KS potential perturbation ($u_{\vec G=0}^{i}(\vec q, \omega)$)  and the dynamic density perturbation ($\rho_{\vec G=0}^{i}(\vec q, \omega)$), which are not standard outputs of currently available LR-TDDFT codes for extended systems (to our best knowledge). Here we show how to circumvent this problem. 

Let us consider the time dependent external perturbation $\delta V_{\rm ext} ~(\vec r, t)=A f(t)\cos(\vec q\cdot \vec r)$ to access the dynamic density response,  where $f(t)$ is a time dependent function, e.g., $f(t)$ can be in the form of a Gaussian envelope \cite{dynamic2}.

In the LRT, since for the considered cosinuoidal potential we have  ${\rm Im}~\mathcal{F}[\delta V_{\rm ext}]=0$ (where $\mathcal{F}[...]$ denotes a Fourier transform), we can use $\chi_{\vec G=0}^{i}(\mathbf{q},\omega)$ to compute ${\rm Re}~ \rho_{\vec G=0}^{i}(\mathbf{q},\omega)$ and ${\rm Im}~ \rho_{\vec G=0}^{i}(\mathbf{q},\omega)$,
\begin{equation}\label{eq:re_chi}
{\rm Re}~{\rho_{\vec G=0}^{i}(\vec q, \omega)}= {\rm Re}~ \chi_{\vec G=0}^{i}(\mathbf{q},\omega)\mathcal{F}[\delta V_{\rm ext}]\ ,    
\end{equation}
\begin{equation}\label{eq:im_chi}
{\rm Im}~{\rho_{\vec G=0}^{i}(\vec q, \omega)}= {\rm Im}~ \chi_{\vec G=0}^{i}(\mathbf{q},\omega)\mathcal{F}[\delta V_{\rm ext}]\ .    
\end{equation}

Eq. (\ref{eq:re_chi}) and Eq. (\ref{eq:im_chi}) define ${\rho_{\vec G=0}^{i}(\vec q, \omega)}$, which is a complex function. 
We combine  ${\rho_{\vec G=0}^{i}(\vec q, \omega)}$ with the KS response function $\chi_{\rm KS, \vec G=0}^{i}(\mathbf{q},\omega)$ to find

\begin{equation}\label{eq:u_G0}
    \begin{split}
    {u_{\vec G=0}^{i}(\vec q, \omega)}&= \frac{\rho_{\vec G=0}^{i}(\vec q, \omega)}{\chi_{\rm KS,\vec G=0}^{i}(\mathbf{q},\omega)}\\
    &= \mathcal{F}[\delta V_{\rm ext}] \frac{\chi_{\vec G=0}^{i}(\mathbf{q},\omega)}{\chi_{\rm KS,\vec G=0}^{i}(\mathbf{q},\omega)}\ ,
    \end{split}
\end{equation}
where $\chi_{\rm KS,\vec G=0}^{i}(\mathbf{q},\omega)$ is computed for each snapshot using the LR-TDDFT result for the macroscopic dynamic dielectric function $\varepsilon^{i,\rm RPA}_{M}(\vec q, \omega)$ in the random phase approximation (RPA) (i.e.,  with zero XC kernel) \cite{moldabekov_lr-tddft_2023},

\begin{equation}\label{eq:macro_ks}
    \chi_{\rm KS,\vec G=0}^{i}(\mathbf{q},\omega)=\frac{1}{v(q)}\left(1-\varepsilon^{i,\rm RPA}_{M}(\vec q, \omega)\right)\ .
\end{equation}

Finally, using Eqs. (\ref{eq:dyn_chi_KS})-(\ref{eq:u_G0}), we find

\begin{equation}
\begin{split}\label{eq:chi_KS_v2}
    \chi_{\rm KS}(\mathbf{q},\omega)&= \frac{\braket{\rho(\vec q, \omega)}_{\vec G=0}}{\braket{ u(\vec q, \omega)}_{\vec G=0}}\\
    &=\frac{\sum_{i=1}^{N_s}{\rho_{\vec G=0}^{i}(\vec q, \omega)}}{\sum_{i=1}^{N_s}{u_{\vec G=0}^{i}(\vec q, \omega)}}\\
        &=\left(\sum_{i}^{N_s}\chi_{\vec G=0}^{i}(\vec q, \omega)\right)\left( \sum_{i=1}^{N_s}\frac{\chi_{\vec G=0}^{i}(\mathbf{q},\omega)}{\chi_{\rm KS,\vec G=0}^{i}(\mathbf{q},\omega)} \right)^{-1},
    \end{split}
\end{equation}
where 
\begin{equation}\label{eq:chi_i}
    \chi_{\vec G=0}^{i}(\mathbf{q},\omega)=\frac{\chi_{\rm KS,\vec G=0}^{i}(\mathbf{q},\omega)}{1-\left[v(q)+K_{\rm xc}^{i}(q)\right]\chi_{\rm KS,\vec G=0}^{i}(\mathbf{q},\omega)}\ .
\end{equation}

Therefore, one needs to compute the KS response function $\chi_{\rm KS,\vec G=0}^{i}(\mathbf{q},\omega)$  using Eq. (\ref{eq:chi_ks_individual}) and the static XC kernel defiend by Eq. (\ref{eq:invert_i}) for each snapshot to find $\chi_{\rm KS}(\mathbf{q},\omega)$ using Eq. (\ref{eq:chi_KS_v2}). LR-TDDFT with zero XC kernel delivers  $\chi_{\rm KS,\vec G=0}^{i}(\mathbf{q},\omega)$ and the direct perturbation method allows one to compute $K_{\rm xc}^{i}(q)$ for a given snapshot.

As it should be in the linear response regime, the parameters of an external perturbation do not enter Eq. (\ref{eq:chi_KS_v2}). 
In the case of a perfect crystal---due to use of a primitive cell   or  a conventional cell---all snapshots are equivalent and we find from  Eq. (\ref{eq:chi_KS_v2})  that $\chi_{\rm KS}(\mathbf{q},\omega)=\chi_{\rm KS, \vec G=0}^{i}(\mathbf{q},\omega)$, again, as it should be for a macroscopic KS response function of crystal. 

In the static limit, Eq. (\ref{eq:chi_KS_v2})  based on the LR-TDDFT gives an equivalent result to Eq. (\ref{eq:chi_KS}) based on the direct perturbation approach.  

To the best of our knowledge,  the formula (\ref{eq:chi_KS_v2})  for the computation of the macroscopic KS response function that has been averaged properly over snapshots  had not been presented in prior works.
We stress that $ \chi_{\rm KS}(\mathbf{q},\omega)\neq \frac{1}{N_s}\sum_{i=1}^{N_s}\chi_{\rm KS,\vec G=0}^{i}(\mathbf{q},\omega)$.

\subsubsection{LR-TDDFT with adiabatic (static) exchange--correlation kernel}

We can now  formulate a consistent \textit{adiabatic (static) approximation} for the macroscopic dynamic density response function
corresponding to properly averaged values of the density and KS potential perturbations,
\begin{eqnarray}\label{eq:chi_adiabatic}
 \chi(\mathbf{q},\omega) = \frac{\chi_{\rm KS}(\mathbf{q},\omega)}{1 -\left[v(q)+K_{\rm xc}(\mathbf{q})\right]\chi_{\rm KS}(\mathbf{q},\omega)}\ ,
\end{eqnarray}
where the dynamic KS response function $\chi_{\rm KS}(\mathbf{q},\omega)$ is given by  Eq. (\ref{eq:chi_KS_v2}) and the static XC kernel is defined by Eq. (\ref{eq:invert}).

The dynamic dielectric function that is consistent with Eq. (\ref{eq:chi_adiabatic}) is defined as

\begin{eqnarray}\label{eq:eps}
\varepsilon^{-1}(\mathbf{q},\omega) =1+v(q)\chi(\mathbf{q},\omega)\ .
\end{eqnarray}

It is clear that
\begin{equation}\label{eq:cor_av_eps}
    \varepsilon(\mathbf{q},\omega) \neq \frac{1}{N_s}\sum_i^{N_s} \varepsilon^{i}(\mathbf{q},\omega)\ ,
\end{equation}
where $\varepsilon^{i}(\mathbf{q},\omega)$ is the macroscopic dynamic dielectric function of an individual snapshot defined as
\begin{eqnarray}\label{eq:eps_i}
\frac{1}{\varepsilon^{i}(\mathbf{q},\omega)}=1+v(q)\chi_{\rm G=0}^{i}(\mathbf{q},\omega)\ .
\end{eqnarray}

Previously, an {adiabatic (static) approximation} for the dynamic density response function  has been formulated 
for a given snapshot of ion positions \cite{moldabekov_lr-tddft_2023}. For the application to  disordered systems,  we have shown in this work how to overcome finite size effects present in  $\chi_{\vec G=0}^{i}(\mathbf{q},\omega)$ due to the snapshot--dependence  by using  the averaged values of the density and KS potential perturbations. 

We stress that inequality~(\ref{eq:cor_av_eps}) holds for other methods such as RT-TDDFT, where the consistent way is to perform averaging on the level of the density response function and then compute the dielectric function.

\section{Simulation details}\label{s:details}

For the computation of the static density response and the static KS response function on the basis of the direct perturbation method, we used the ABINIT package \cite{Gonze2020, Romero2020, Gonze2016, Gonze2009, Gonze2005, Gonze2002} with the PBE \cite{PBE} XC functional.
The simulations are performed for warm dense hydrogen at $r_s=2$ and $r_s=4$, where $r_s=\left(4\pi n_0/3\right)^{-1/3}$ is the mean-inter particle distance and $n_0$ is the mean number density of electrons (protons).
In Ref. \cite{Moldabekov_dft_kernel}, it was shown that KS-DFT simulations of hydrogen at the considered WDM parameters provide an accurate description of the static density response  by comparing with available quantum Monte Carlo data~\cite{Bohme_PRL_2022,Bohme_PRE_2022}.

We consider ion snapshots with $N=14$ and $N=112$ particles. In the case of  $N=14$ particles, results are computed for $N_s=10$ different snapshots generated by segmenting a larger ionic configuration that has been obtained from a thermal KS-DFT based molecular dynamics simulations as it is described in Ref. \cite{Fiedler_PRR_2022}. We consider temperatures corresponding to partial electron degeneracy with $\theta=T/T_F=1$, where $T_F$ is the Fermi temperature of free electrons. At $r_s=2$ ($r_s=4$), we have $T\simeq 12.528~{\rm eV}$ ($T\simeq 3.132~{\rm eV}$). For the LR-TDDFT calculations of the macroscopic KS response function, we used the GPAW code \cite{LRT_GPAW2}.

For the direct perturbation approach based KS-DFT calculation with $N=14$ ($N=112$) particles in a snapshot,  we used $N_b=280$ ($N_b=2000$) bands in the main simulation cell.  For $N=14$ particles, the k-points sampling was set to $10\times10\times10$  with an energy cutoff of $30~{\rm Ha}$. Additionally, for $r_s=4$, we present results for $N=378$ particles with $N_b=7600$ bands. For  $N=112$ and $N=378$ particles, the k-points sampling was set to $2\times2\times2$  with the energy cutoff $30~{\rm Ha}$.
The box size for $N=14$ ($N=112$) is  $L=7.77~{\rm Bohr}$ ($L=15.541~{\rm Bohr}$), which is defined by the relation $n_0 L^3=N$.
The amplitude of the external perturbation is set $A=0.01$ (in Hartree) . It was shown to be within the LRT domain in Ref. \cite{Bohme_PRL_2022}. The results are presented in Hartree atomic units.

The convergence of KS-DFT simulations with respect to  \textit{k}-point grid, energy cutoff,  and the number of bands have been studied in \cite{Moldabekov_JCP_2021, hybrid_results, moldabekov_lr-tddft_2023} for the uniform electron gas (UEG), which has properties similar to fully ionized hydrogen, and warm dense hydrogen \cite{Moldabekov_dft_kernel, moldabekov_lr-tddft_2023}. 
To ensure the convergence of the presented results, in this work, we set same or better computation parameters. 
For example, the convergence of KS-DFT calculations is tested by reproducing an exact Lindhard response function in the thermodynamic limit in the case of UEG \cite{moldabekov_lr-tddft_2023}.  For warm dense hydrogen, the high-quality of the KS-DFT simulations at considered parameters is tested on the example of LDA XC functional by comparing to the QMC data for the density response   \cite{Moldabekov_dft_kernel, moldabekov_lr-tddft_2023}.

We set $\vec q$ along the z-axis. We consider the response of the system along $\vec q$  and drop the vector notation for simplicity. 
The density and KS potential perturbation values are averaged along the $x$ and $y$ axes.
The perturbation wavenumbers are defined as $q=j\times q_{\rm min}$, where $q_{\rm min}=2\pi/L$ and $j$ denotes a positive integer number.
For $N=14$, we have $q_{\rm min}^{N=14}\simeq 0.84268~q_F$, for $N=112$ we have $q_{\rm min}^{N=112}\simeq 0.42134~q_F$, and for $N=378$ we have $q_{\rm min}^{N=378}\simeq 0.280894~q_F$.
Since  $q_{\rm min}^{N=14}=2q_{\rm min}^{N=112}$ and $q_{\rm min}^{N=14}=3q_{\rm min}^{N=378}$, we can compare results computed using $N=14$, $N=112$, and $N=378$ particles.

To demonstrate the application of Eq. (\ref{eq:chi_KS_v2}), LR-TDDFT calculations of the macroscopic dynamic KS response function are performed for  $r_s=2$ using $N_s=10$ different snapshot with $N=14$ electrons (protons) in each, $10\times10\times10$ k-points, a cutoff in the dielectric function of $100~{\rm eV}$ and a broadening parameter  $\eta=0.2$. 
 
\section{Results and discussions}\label{s:results}

We first present the results for the static density response, static KS response, and static XC kernel computed using the direct perturbation approach and theory presented in Sec. \ref{s:theory_A}. After that, we demonstrate the application of the averaging scheme for the dynamic density response function presented in Sec. \ref{s:theory_B}.

\subsection{Static density response and XC kernel}

We consider warm dense hydrogen at $r_s=2$ and $r_s=4$. The former corresponds to a characteristic metallic density with a high ionization degree and the latter to a partially ionized dense gaseous state consisting of ions and neutral atoms \cite{Bohme_PRL_2022, Bohme_PRE_2022, Militzer_PRE_2001}. Therefore,  we have stronger electron-ion coupling at $r_s=4$ than at $r_s=2$ and, correspondingly, a more pronounced degree of inhomogeneity in the electron density for a given snapshot. 

\begin{figure}\centering
\includegraphics[width=0.45\textwidth]{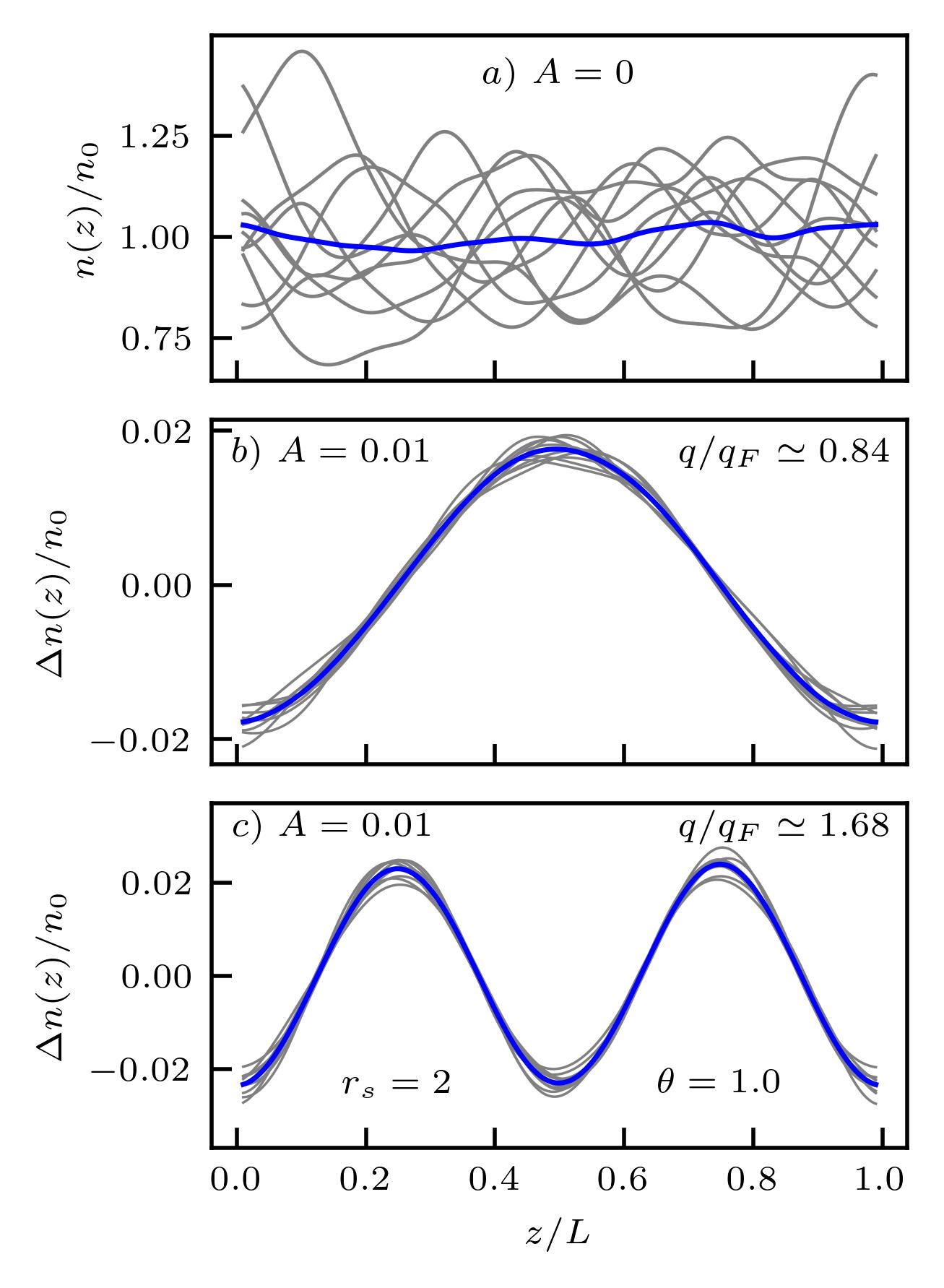}
\caption{\label{fig:den_rs2} a) Density distribution along the z axis of the unperturbed system, b) density perturbation for $A=0.01$ and $q\simeq 0.84q_F$, and  c) density perturbation for $A=0.01$ and $q\simeq 1.68q_F$. The results are computed for  $r_s=2$ and $\theta=1$. 
}
\end{figure} 

 \begin{figure}\centering
\includegraphics[width=0.45\textwidth]{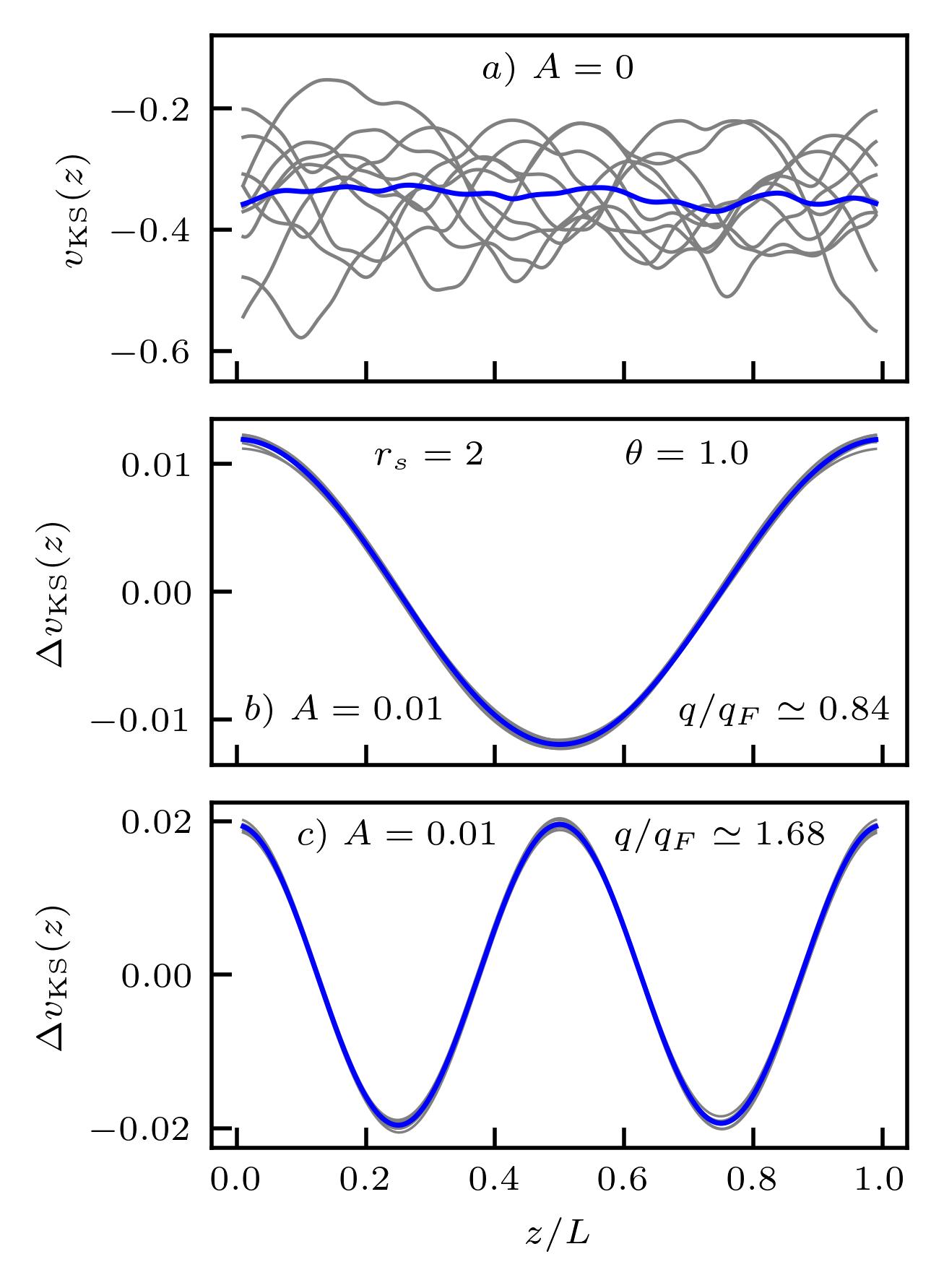}
\caption{\label{fig:pot_rs2} a) KS potential distribution along the z axis of the unperturbed system, b) KS potential perturbation for $A=0.01$ and $q\simeq 0.84q_F$, and  c) KS potential  perturbation for $A=0.01$ and $q\simeq 1.68q_F$. The results are computed for  $r_s=2$ and $\theta=1$. 
}
\end{figure}

\subsubsection{Hydrogen at metallic density, $r_s=2$}

In Fig. \ref{fig:den_rs2}, we show the unperturbed electron density as well as the density perturbations  due to an external harmonic field for 10 different snapshots (grey curves); the solid blue lines depict the corresponding mean values. From Fig. \ref{fig:den_rs2} a), we clearly see that the unperturbed densities are  inhomogeneous and that averaging over snapshots leads to a homogeneous density profile. In Fig. \ref{fig:den_rs2} b) and Fig. \ref{fig:den_rs2} c), we present results for the perturbation wavenumbers $q=q_{\rm min}^{N=14}\simeq 0.84268~q_F$  and $q=2q_{\rm min}^{N=14}$. We observe that averaging leads to a cancellation of the small deviations from the cosinuoidal shape of the density perturbation following the shape of the external perturbation.

\begin{figure*}
\includegraphics[width=\linewidth]{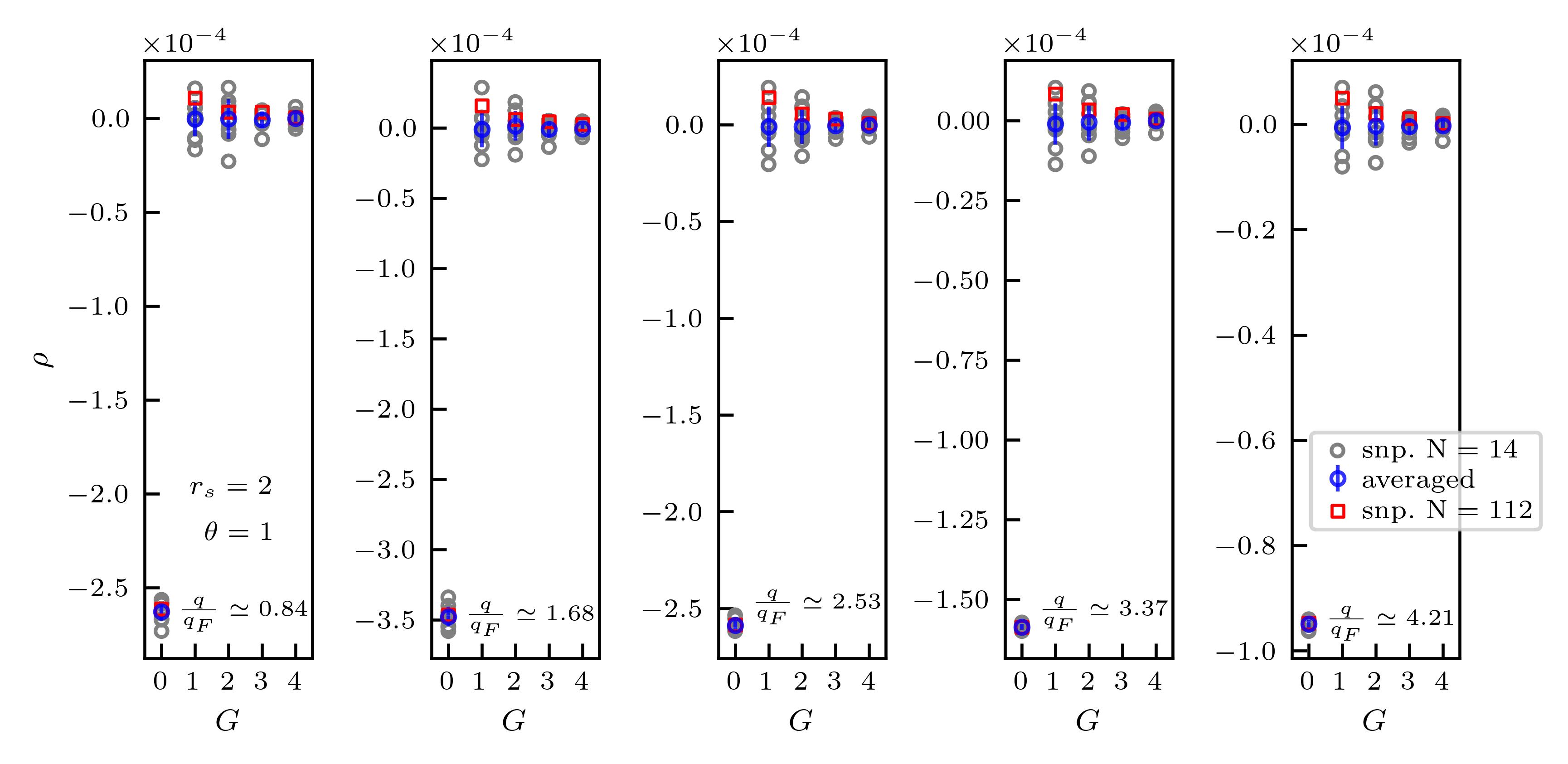}
\caption{\label{fig:rho_rs2}  Contributions to the total density change from density perturbation values at different wave numbers for different snapshots (the grey circles are for 14 particles and red symbols are for 112 particles), and for the averaged values over 10 snapshots with 14 particles (blue circles) in warm dense hydrogen at $r_s=2$ and $\theta=1$.
The wave number $q$ corresponds to the wavenumber of the external perturbation. The $\vec G$ is along the z-axis and in units of $2\pi/L$.
}
\end{figure*}

\begin{figure*}
\includegraphics[width=\linewidth]{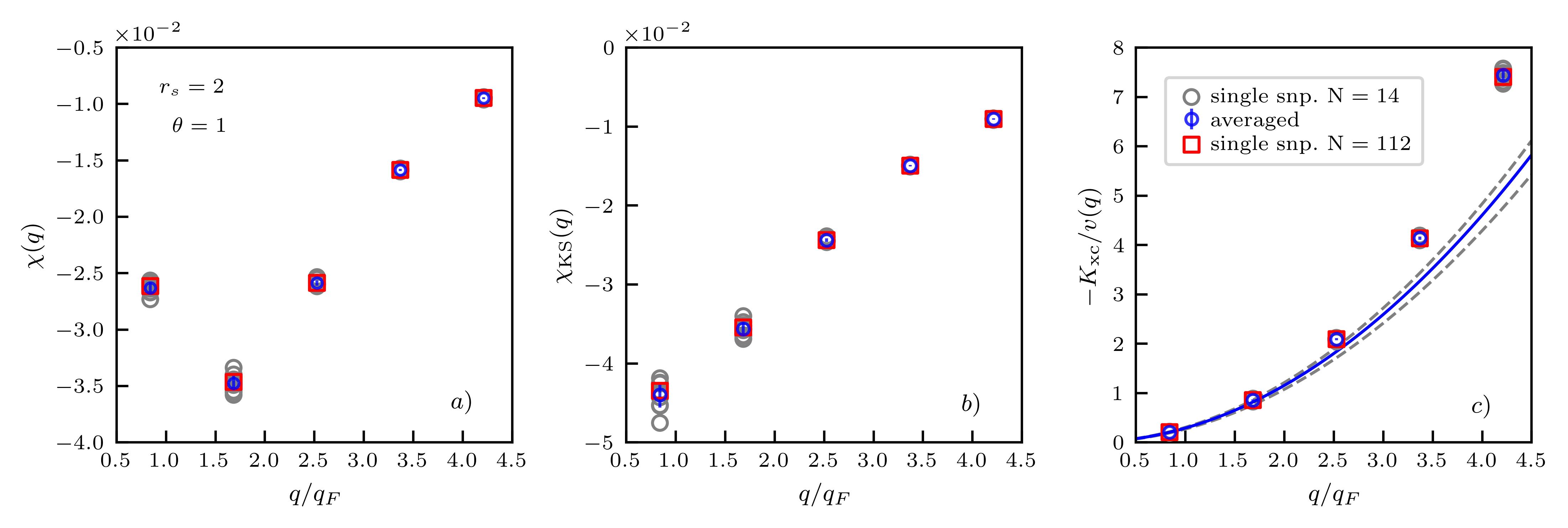}
\caption{\label{fig:chi_rs2}  a) Total static density response function, b) static KS response function, and c) static XC kernel of warm dense hydrogen at $r_s=2$ and $\theta=1$.
Grey circles are for 14 particles and red squares are for 112 particles. Blue circles are for the averaged values over 10 snapshots with 14 particles.
}
\end{figure*}

In Fig. \ref{fig:pot_rs2}, we present results for the KS potential of the unperturbed and perturbed systems.
Similarly to the density distribution, the KS potential profile is inhomogeneous for individual snapshots and tends to  the homogeneous  distribution upon averaging over snapshots (see Fig. \ref{fig:pot_rs2} a)). In the case of the perturbed system, the averaged value of the KS potential perturbation, presented in Fig. \ref{fig:pot_rs2} b) and Fig. \ref{fig:pot_rs2} c), closely follows  the external perturbation. We note that the deviation of the KS potential perturbation for individual snapshots from  the cosinuoidal shape is less pronounced compared to the density perturbation profiles at the same parameters (cf. Fig. \ref{fig:den_rs2} b) and Fig. \ref{fig:den_rs2} c)).

\begin{figure*}
\includegraphics[width=\linewidth]{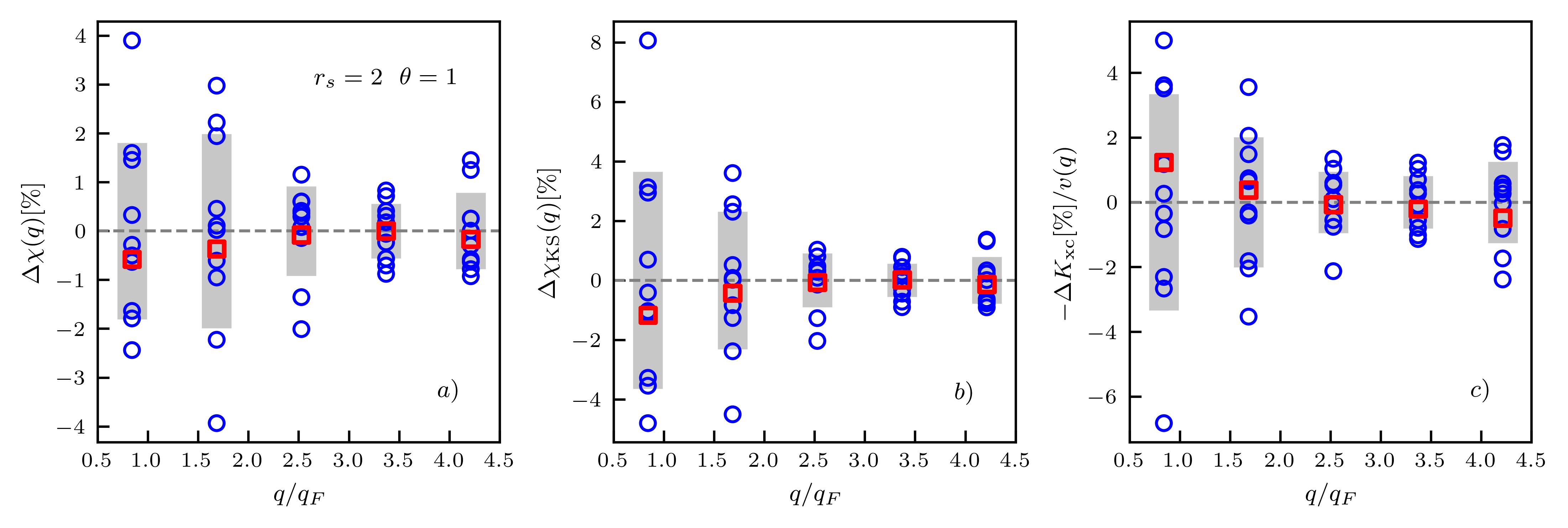}
\caption{\label{fig:diff_chi_rs2} Difference of   a) the total static density response function, b) the static KS response function, and  c) the static XC kernel computed for a single snapshot from the corresponding averaged values at $r_s=2$ and $\theta=1$.
Colored circles are for the snapshots with 14 particles and  red symbols are for the snapshot with  112 particles.
}
\end{figure*}

From the results presented in Fig. \ref{fig:den_rs2} and Fig. \ref{fig:pot_rs2}, it is clear that averaged values of the density perturbation and KS potential perturbation have a cosinuoidal shape with the wavenumber $q$ being equal to that of the external harmonic field.
To further confirm this observation, we show in Fig. \ref{fig:rho_rs2} the contributions to the total density change from density perturbation values $\rho_{G}(q)$ (computed using Eq. (\ref{eq:expansion})) at different wavenumbers $G$ and $q$.  
In Fig. \ref{fig:rho_rs2}, we present results for 10 different snapshots of $N=14$ particles, for the corresponding averaged values, and for one snapshot with $N=112$ particles.
Additionally, we show corresponding standard deviations of a single snapshot, which are depicted as ``error bars''.
From Fig. \ref{fig:rho_rs2}, we see that $\rho_{G\neq 0}^{i}(q)$ values for different snapshots have different signs, and have magnitudes much less than that of   $\rho_{G= 0}^{i}(q)$. In contrast, all $\rho_{G= 0}^{i}(q)$ values for the considered snapshots have the same sign and relatively close values; this is shown numerically below considering density response functions. 
As a result, we have  $\braket{\rho(q)}_{G=0} \gg \braket{\rho(q)}_{G\neq0 }$  for the mean values and, compared to the contribution at $G=0$, one can set $\braket{\rho(q)}_{G\neq0 }\approx 0$. Furthermore, we see that the averaged value  $\braket{\rho(q)}_{G= 0}$ computed for $N=14$ particles is in close agreement with $\rho_{G=0}^{i}(q)$ computed for $N=112$ particles. In contrast, there are significant disagreements between $\rho_{G\neq 0}^{i}(q)$ for a snapshot with $N=112$ particles and $\braket{\rho(q)}_{G\neq0}$ computed by averaging over 10 snapshots with $N=14$ particles in each. Additionally, we observe that $\rho_{G\neq 0}^{i}(q)$ for the snapshot with $N=112$ particles has magnitudes comparable with $\rho_{G\neq 0}^{i}(q)$ for a single snapshot with $N=14$ particles.

In Fig. \ref{fig:chi_rs2}, we show the results for a) the static density response function, b) the static KS response function, and c) the static XC kernel of warm dense hydrogen at $r_s=2$ and $\theta=1$. In Fig. \ref{fig:chi_rs2} a), the averaged value of the static density response function $\chi(q)$ is computed  using Eq. (\ref{eq:sum_chi}) and the static density response function for an individual snapshot $\chi_{G=0}^{i}(q)$ is computed using Eq. (\ref{eq:delta_n_LRT_G}). As one can see from Fig. \ref{fig:chi_rs2} a), the $\chi(q)$ computed for $N=14$ particles is in good agreement with $\chi_{G=0}^{i}(q)$  computed for $N=112$ particles. The standard deviations for snapshots with $N=14$ particles are also shown. 

In Fig. \ref{fig:chi_rs2} b), the macroscopic static KS response function   $\chi_{{\rm KS}}(q)$ characterizing the density response to the change in the KS potential on average is computed using Eq. (\ref{eq:chi_KS}) (presented with corresponding standard deviations of a single snapshot), and the static KS response function for an individual snapshot $\chi_{{\rm KS}, G=0}^{i}(q)$ is computed using Eq. (\ref{eq:chi_ks_individual}). From Fig. \ref{fig:chi_rs2} b), we see that  
 $\chi_{{\rm KS}}(q)$ computed for $N=14$ particles is in close agreement with $\chi_{{\rm KS}, G=0}^{i}(q)$ calculated for a snapshot with $N=112$ particles. In contrast,  the $\chi_{{\rm KS}, G=0}^{i}(q)$ values obtained using snapshots with $N=14$ particles have visible disagreements with the  results for $\chi_{{\rm KS}}(q)$.

To analyze the data for the static XC kernel $K_{\rm xc}(q)$, we use the so-called local field correction~\cite{kugler1} $-K_{\rm xc}(q)/v(q)$, which is commonly used for the study of the dielectric properties of homogeneous systems such as quantum Fermi liquids \cite{quantum_theory} and, in particular,  uniform electron gas \cite{review}. We note that the  local field correction is not related to the term ``local field effects'' used in the context of LR-TDDFT to describe the density inhomogeneity induced by the ions.
In Fig. \ref{fig:chi_rs2} c), we present the data for the static XC kernel $K_{\rm xc}(q)$ computed using Eq. (\ref{eq:invert}). The results for $K_{\rm xc}(q)$ and the corresponding standard deviations are based on the data generated for $N_s=10$ snapshots with $N=14$ particles in each of them.
We compare the $K_{\rm xc}(q)$  with the $K^{i}_{\rm xc}(q)$  values  computed using Eq. (\ref{eq:invert_i}) for each snapshot separately with $G=0$. Additionally, we compare with the $K^{i}_{\rm xc}(q)$ values calculated for one snapshot with $N=112$ particles. 
Additionally, we plot a quadratic dependence accurately describing  the local field correction $-K_{\rm xc}(q)/v(q)$ at $q\lesssim 1.5q_F$. The solid (blue) line is obtained using the $K_{\rm xc}(q)$ value at $q<q_F$ and the dashed (grey) lines are defined by the smallest and largest values of the $K^{i}_{\rm xc}(q)$ (at $q<q_F$) among considered snapshots. From Fig. \ref{fig:chi_rs2} c), we observe a close agreement between $K_{\rm xc}(q)$ based on averaged quantities and  $K^{i}_{\rm xc}(q)$ computed for one snapshot with $N=112$ particles.  From Fig. \ref{fig:chi_rs2}, we see that  the local field corrections $-K_{\rm xc}(q)/v(q)$  and $-K_{\rm xc}^{i}(q)/v(q)$  follow a quadratic behavior at $q\lesssim 1.5q_F$ and show faster increase than the quadratic dependence with the increase in the wavenumber at $q >2q_F$.

To further quantify the difference between the results for individual snapshots and for the averaged values of  the considered density response characteristics, we plot the relative deviation of $\chi_{G=0}^{i}(q)$ from $\chi(q)$ computed as
\begin{equation}\label{eq:d_chi_chi}
    \Delta \chi (q)=\frac{\chi_{G=0}^{i}(q)-\chi(q)}{\chi(q)}\times 100\%
\end{equation}
  in  Fig. \ref{fig:diff_chi_rs2}a).
Further, we show the relative deviation of  $\chi_{{\rm KS},G=0}^{i}(q)$ from $\chi_{\rm KS}(q)$,
\begin{equation}\label{eq:d_chiks_chiks}
    \Delta \chi_{\rm KS} (q)=\frac{\chi_{{\rm KS}, G=0}^{i}(q)-\chi_{\rm KS}(q)}{\chi_{\rm KS}(q)}\times 100\%\ ,
\end{equation}
in  Fig. \ref{fig:diff_chi_rs2}b), the relative deviation of  $K_{\rm xc}^{i}(q)$ from $K_{\rm xc}(q)$,
\begin{equation}\label{eq:d_kxc_kxc}
    \Delta K_{\rm xc}(q)=\frac{K_{\rm xc}^{i}(q)-K_{\rm xc}(q)}{K_{\rm xc}(q)}\times 100\%\ ,
\end{equation}
in Fig. \ref{fig:diff_chi_rs2}c).
The $\Delta \chi (q)$, $\Delta \chi_{\rm KS} (q)$, and $\Delta K_{\rm xc}(q)$ values are shown as blue circles,  grey areas represent the standard deviations (of a single snapshot) estimated using $N_s=10$ snapshots of $N=14$ particles, and red squares depict data computed comparing  the results for one snapshot with $N=112$ particles to the averaged values based on 10 snapshots with $N=14$ particles. 
From Fig. \ref{fig:diff_chi_rs2}, we see that the disagreement between $\chi (q)$ computed by averaging over snapshots with $N=14$ particles and  $\chi_{G=0}^{i}(q)$ computed for a single snapshot with $N=112$ particles is less than $1\%$. 
For $\chi_{\rm KS} (q)$ and $K_{\rm xc}(q)$, the  disagreement with $\chi_{{\rm KS}, G=0}^{i}(q)$ and $K_{\rm xc}^{i}(q)$ of the snapshot with $N=112$ particles, is less than about $1.5\%$. These deviations for the averaged values with $N=14$ particles and the values obtained for a single snapshot with $N=112$ particles are significantly smaller than the standard deviations evaluated using $N_s=10$ snapshots of $N=14$ particles. 
In contrast, results for $\chi_{G=0}^{i}(q)$ , $\chi_{{\rm KS}, G=0}^{i}(q)$ and $K_{\rm xc}^{i}(q)$ for the snapshots with $N=14$ particles deviate from the corresponding averaged values by up to about  $4\%$, $8\%$, and $7\%$, respectively.

 \begin{figure}\centering
\includegraphics[width=0.45\textwidth]{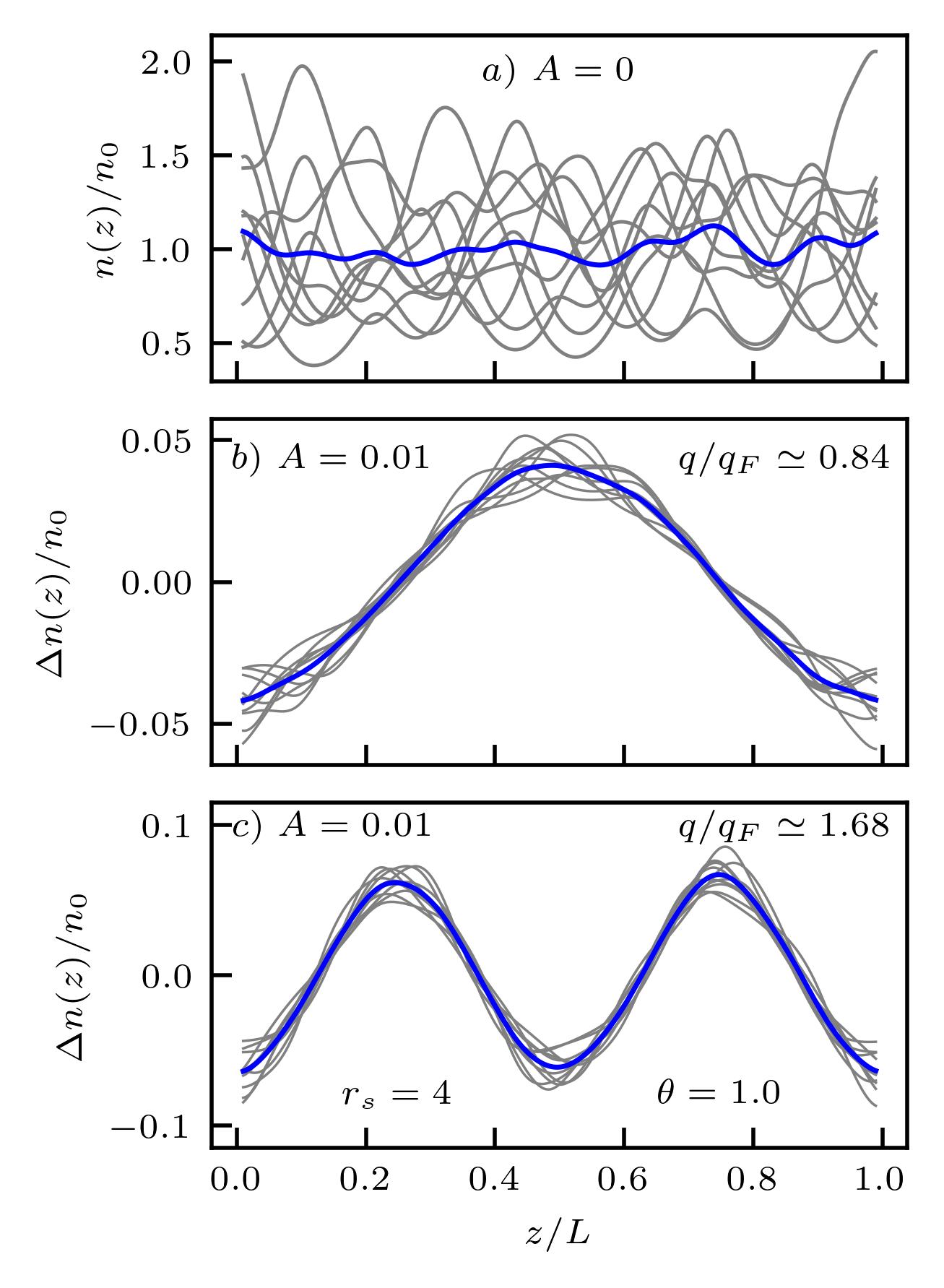}
\caption{\label{fig:den_rs4} a) Density distribution along the z axis of the unperturbed system, b) density perturbation at $A=0.01$ and $q\simeq 0.84q_F$, and  c) density perturbation at $A=0.01$ and $q\simeq 1.68q_F$. The results are computed for  $r_s=4$ and $\theta=1$. 
}
\end{figure} 

 \begin{figure}\centering
\includegraphics[width=0.45\textwidth]{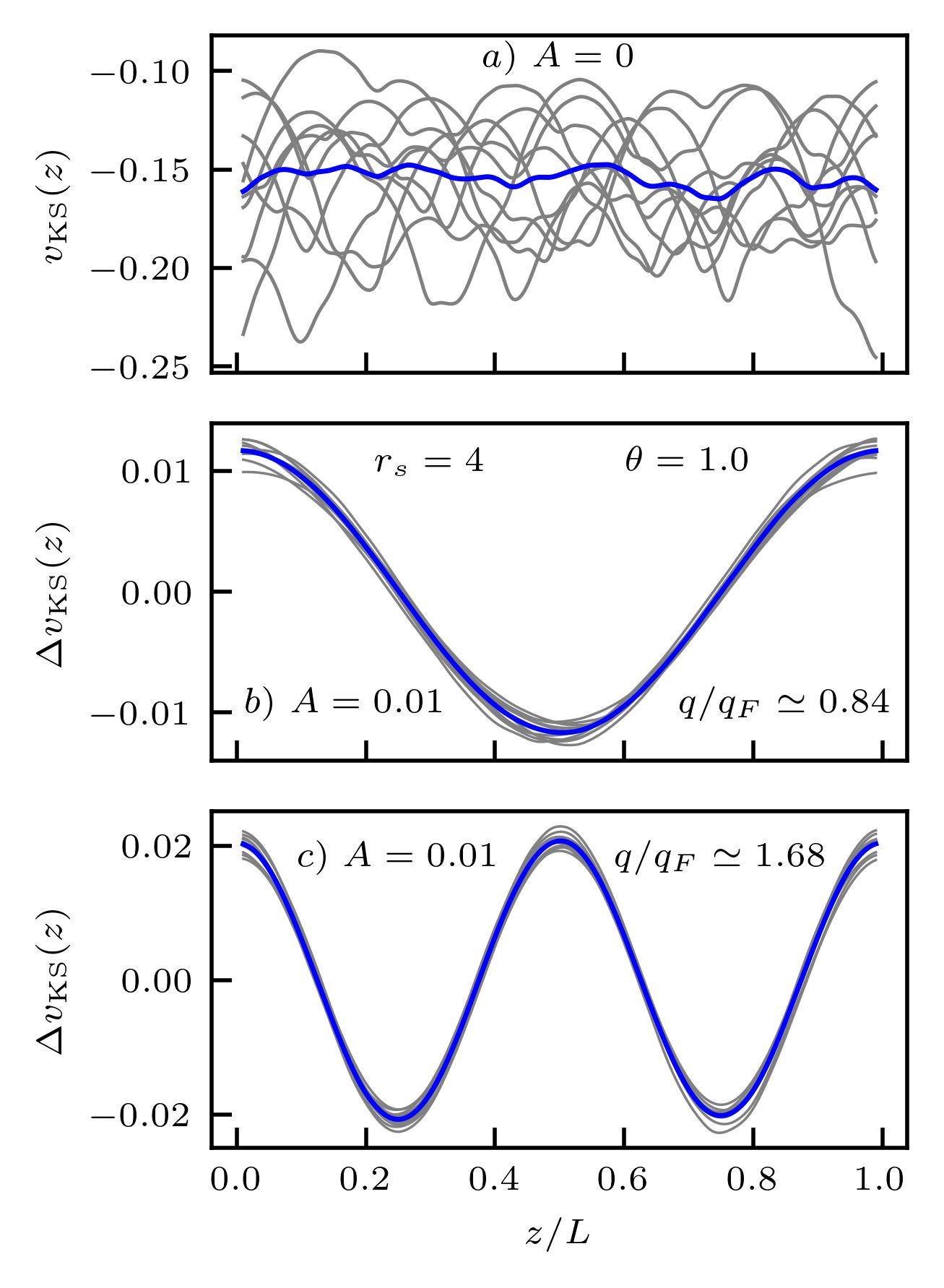}
\caption{\label{fig:pot_rs4} a) KS potential distribution along the z axis of the unperturbed system, b) KS potential perturbation at $A=0.01$ and $q\simeq 0.84q_F$, and  c) KS potential  perturbation at $A=0.01$ and $q\simeq 1.68q_F$. The results are computed for  $r_s=4$ and $\theta=1$. 
}
\end{figure}

\subsubsection{Partially ionized dense hydrogen, $r_s=4$}

At $r_s=4$, we have a stronger coupling between electrons and ions compared to the case with $r_s=2$.
This means that electrons are localised around ions to a larger degree. This can be seen from Fig. \ref{fig:den_rs4} a), where  density profiles for $N_s=10$ different snapshots with $N=14$ particles (solid grey lines) are shown  for the unperturbed dense hydrogen gas at $r_s=4$ and $\theta=1$ ($T\simeq 3.132~{\rm eV}$). From Fig. \ref{fig:den_rs4} a), we see that the density values deviate from the mean density $n_0$ by up to $100\%$. Nevertheless, the averaged value over snapshots of the equilibrium density is homogeneous (solid blue line)  due to the disordered structure at the considered parameters.
In the case of the perturbed system presented in Fig. \ref{fig:den_rs4}b) and Fig. \ref{fig:den_rs4}c), the density averaging over 10 snapshots effectively eliminates the deviations from the cosinuoidal profile in the density perturbation. This is also demonstrated for snapshots with 112 particles in the Appendix B.

\begin{figure*}
\includegraphics[width=\linewidth]{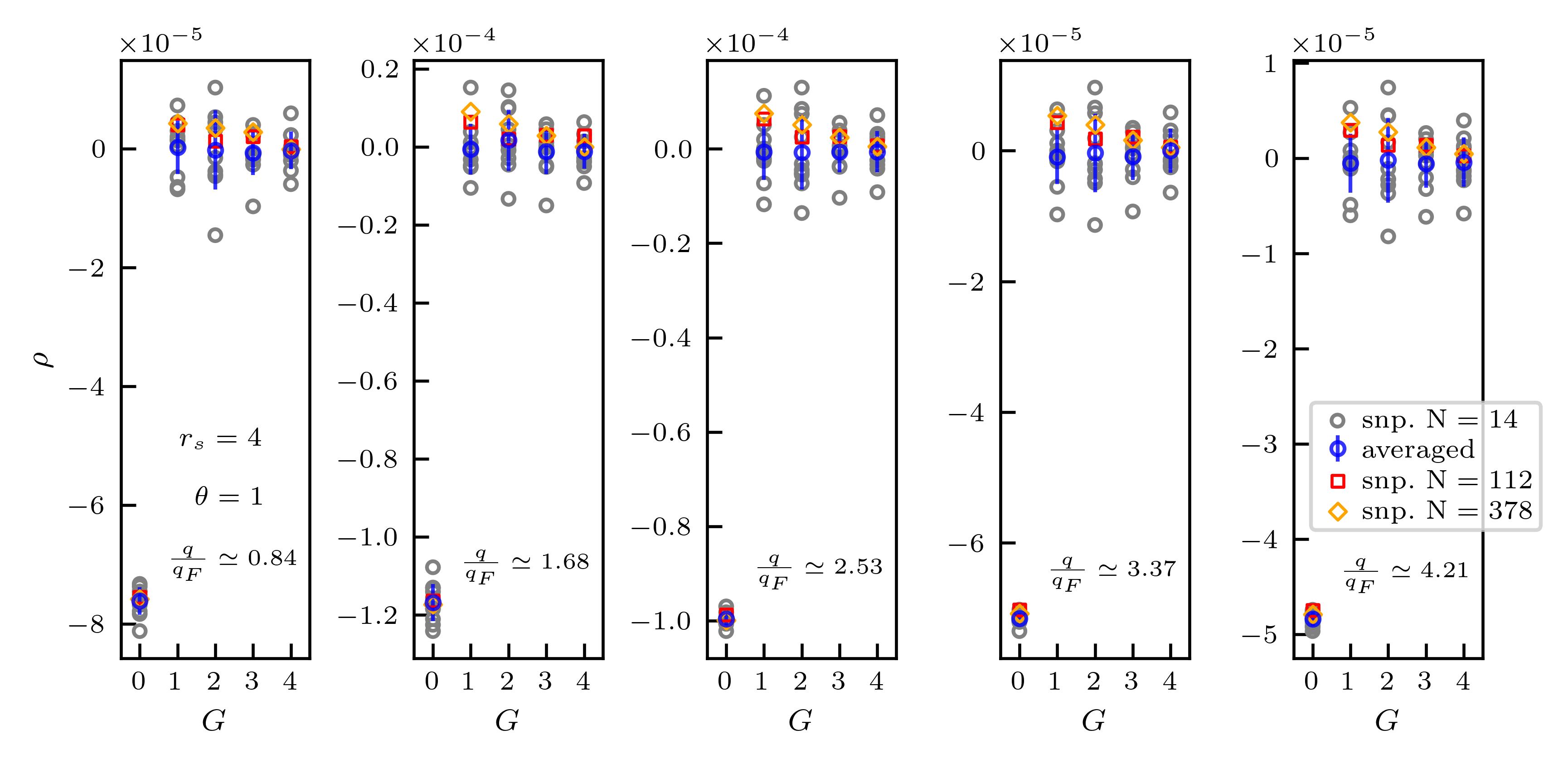}
\caption{\label{fig:rho_rs4}  Contributions to the total density change from the density perturbation values at different wavenumbers for different snapshots (the grey circles are for 14 particles and red (orange) symbols are for 112 (378) particles), and for the averaged values over 10 snapshots with 14 particles (blue circles) in warm dense hydrogen at $r_s=4$ and $\theta=1$.
The wave number $q$ corresponds to the wavenumber of the external perturbation.  The $\vec G$ is along the z-axis and in units of $2\pi/L$.
}
\end{figure*}

\begin{figure*}
\includegraphics[width=\linewidth]{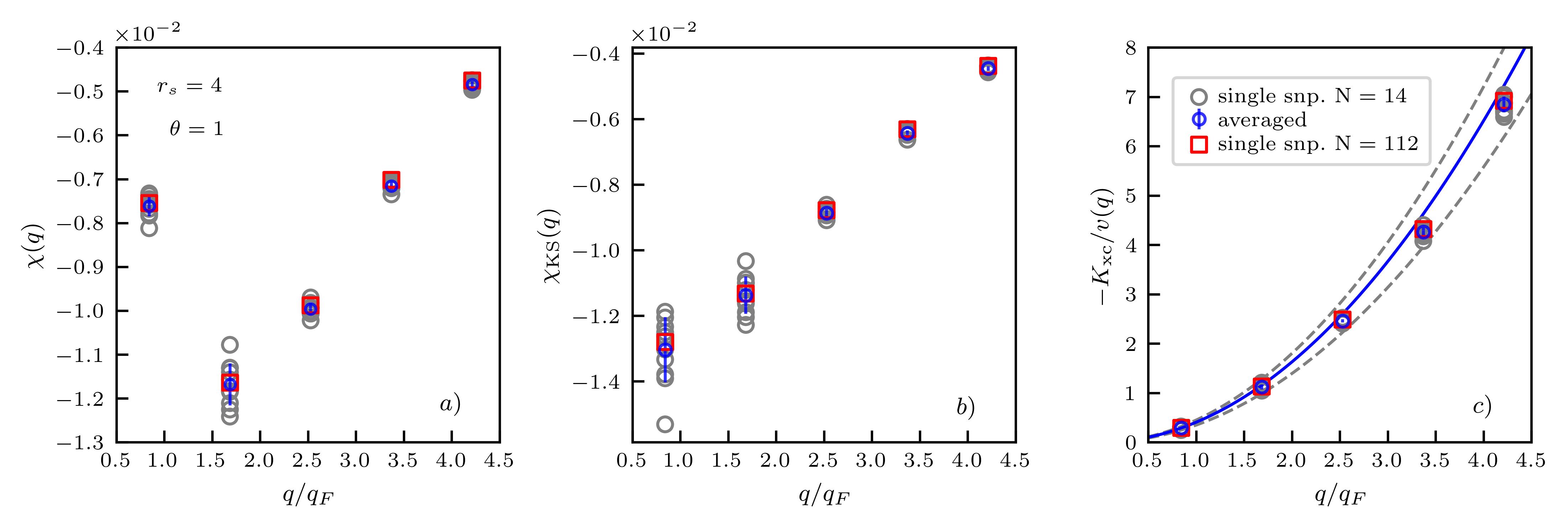}
\caption{\label{fig:chi_rs4}  a) Total static density response function, b) static KS response function, and c) static XC kernel of warm dense hydrogen at $r_s=4$ and $\theta=1$.
Grey circles are for 14 particles and red squares are for 112 particles. Blue circles are for the averaged values over 10 snapshots with 14 particles.
}
\end{figure*}

\begin{figure*}
\includegraphics[width=\linewidth]{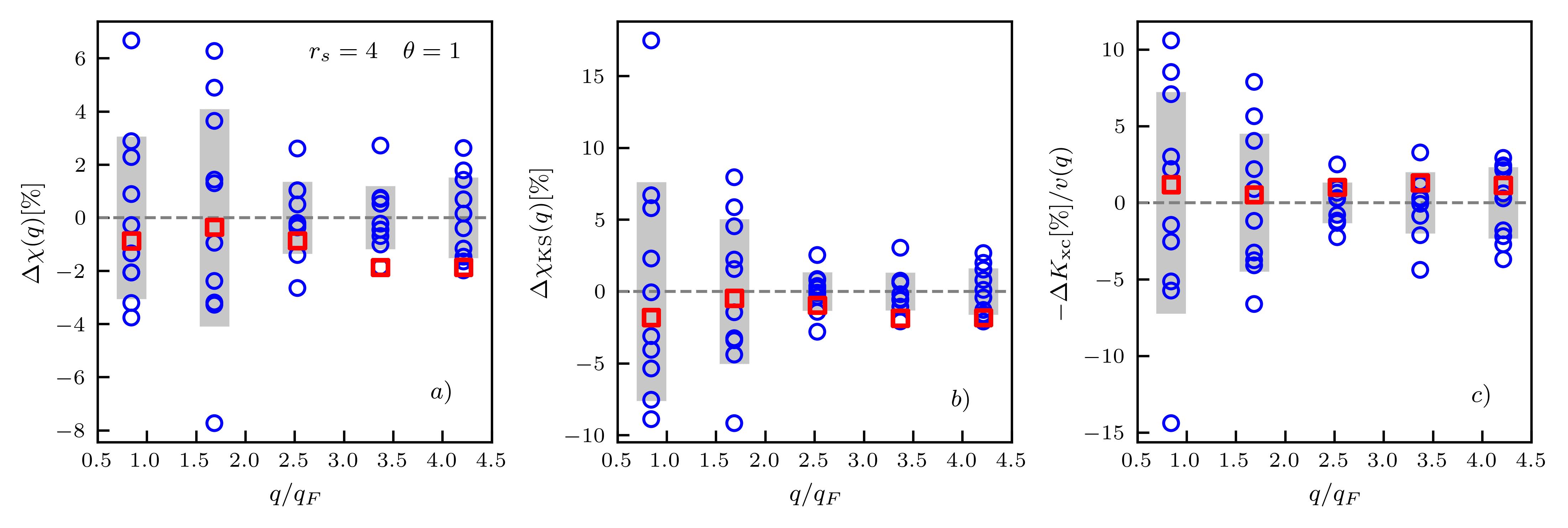}
\caption{\label{fig:diff_chi_rs4} Difference of   a) the total static density response function, b) the static KS response function, and c) the static XC kernel computed for a single snapshot from the corresponding averaged values at $r_s=2$ and $\theta=1$.
Colored circles are for the snapshots with 14 particles and  red symbols are for a snapshot with  112 particles.
}
\end{figure*}

In Fig. \ref{fig:pot_rs4}, we show the KS potential profiles for the unperturbed system and for the perturbed system with $r_s=4$ and $\theta=1$. We see that the KS potential for individual snapshots (with $N=14$ particles) is strongly inhomogeneous.
These inhomogeneities are smoothed out after averaging over snapshots and the KS potential distribution becomes nearly uniform due to disordered structure of the system. In the perturbed systems shown in Fig. \ref{fig:pot_rs4}b) and Fig. \ref{fig:pot_rs4}c), the KS potential distribution follows the external perturbation. The deviations from the cosinuoidal distribution are visible for individual snapshots. These deviations are diminished significantly by averaging over snapshots. 

To further demonstrate that only a single Fourier component at the wavenumber of the external harmonic perturbation remains after averaging  of the density perturbations over snapshots, we present the density perturbation values $\rho_{G}(q)$ (as defined in Eq. (\ref{eq:expansion})) at different $G$ and $q$ values in Fig. ~\ref{fig:rho_rs4}. We find that the $\rho_{G\neq0}(q)$ components have significant contributions for individual snapshots, but with different signs. The latter leads to the mutual cancellation of the $\rho_{G\neq0}(q)$ values from different snapshots after averaging.   In the Appendix B, we also demonstrate it for snapshots with 112 particles.
In contrast, the $\rho_{G=0}(q)$  components of different snapshots have the same sign and after averaging over snapshots we have 
$\braket{\rho(q)}_{G=0}\gg \braket{\rho (q)}_{G\neq0}$, and, comparing with $\braket{\rho(q)}_{G=0}$, one can safely neglect $\braket{\rho(q)}_{G\neq0}$.  Additionally, we observe that, in general, the $\braket{\rho(q)}_{G=0}$ values computed by averaging over $N_s=10$ snapshots of $N=14$ particles are in a good agreement with the  ${\rho(q)}_{G=0}^{i}$ values obtained using a single snapshot of $N=112$ or $N=378$ particles.

\begin{figure*}
\includegraphics[width=\linewidth]{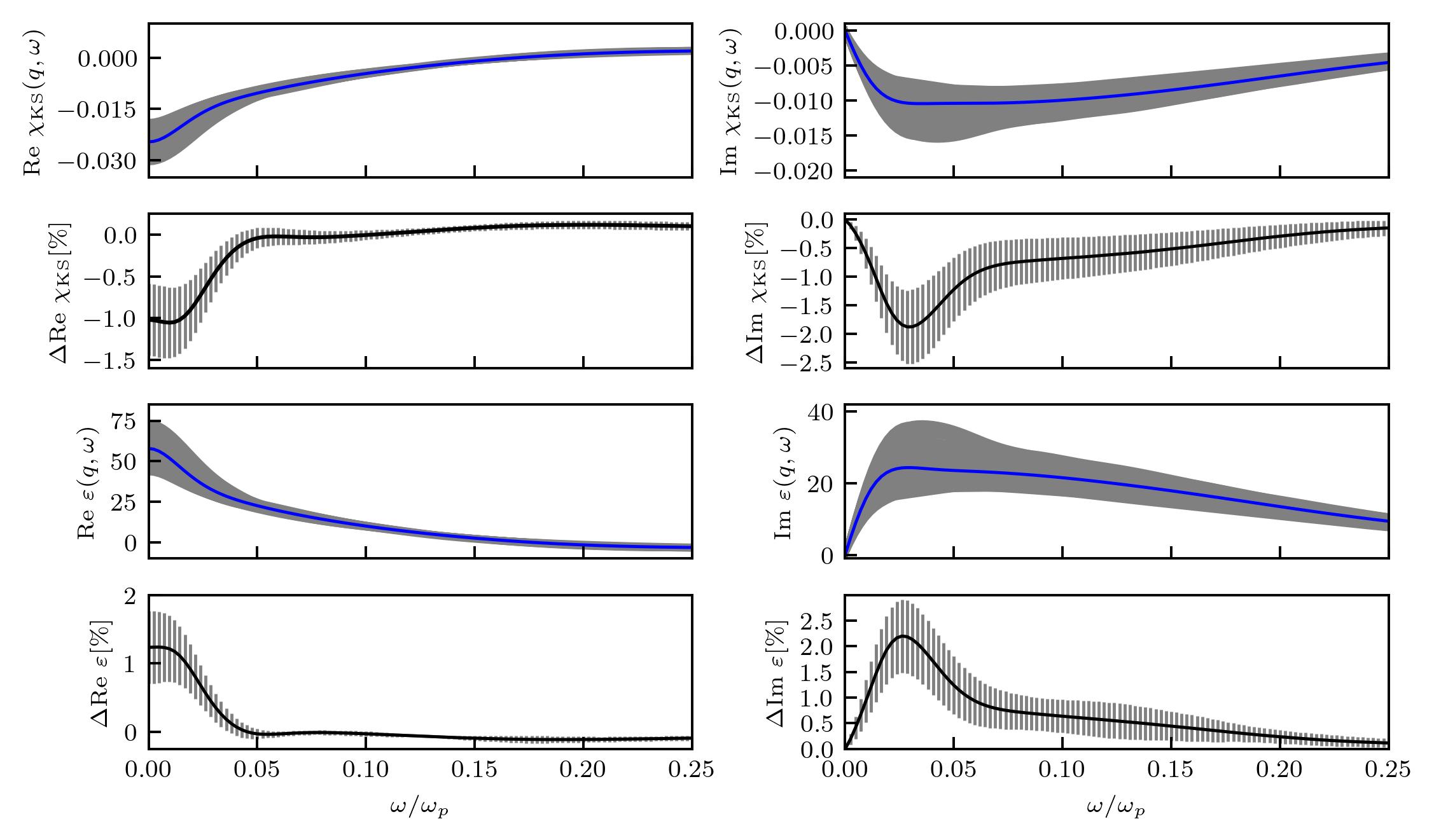}
\caption{\label{fig:eps2}  First row: real and imaginary part of the macroscopic KS response function. 
Second row:  difference between the results for the KS response function computed using different averaging methods. 
Third row: real and imaginary  part of the dielectric function.
Bottom row:  difference between the results for the macroscopic dielectric function computed using different averaging methods. 
The results for a given snapshot are presented by grey lines (with 14 particles), and the averaged values over 250 snapshots  are presented by blue lines. The results  are computed for warm dense hydrogen at $r_s=2$ and $\theta=1$ for $q/q_F\simeq 0.084$.  The standard error is depicted by vertical gray lines in the  second and bottom rows.
}
\end{figure*}

Next, we show in Fig. \ref{fig:chi_rs4} results for a) the density response functions, b) the KS response functions, and for c) the static XC kernels computed for $N_s=10$ snapshots of $N=14$ particles, averaged values over these snapshots, and for a single snapshot with $N=112$ particles. From Fig. \ref{fig:chi_rs4}, we see that the averaged values are in good agreement with the data computed using one snapshot of $N=112$ particles. 
In contrast,  we find that $\chi_{G=0}^{i}(q)$ and $\chi_{{\rm KS}, G=0}^{i}(q)$  of the snapshots with $N=14$ particles deviate significantly from the corresponding averaged values at $q\lesssim 3 q_F$. 
We observe similar trends for the static XC kernel at $q\gtrsim 3q_F$.

 In Fig. \ref{fig:chi_rs4} c), we also show a quadratic approximation for the local field correction $-K_{\rm xc}(q)/v(q)$. The solid (blue) line is computed using the $K_{\rm xc}(q)$ value at $q<q_F$ and the dashed (grey) lines correspond to the smallest and largest values of the $K^{i}_{\rm xc}(q)$ (at $q<q_F$) among the considered snapshots. From Fig. \ref{fig:chi_rs4} c), one can see that the XC kernel describing the averaged value is very well described by the quadratic curve at all considered wavenumbers. For individual snapshots, the XC kernel can increase faster or slower than quadratic upon increasing the wavenumber depending on the configuration of ions.
This is in contrast to the behavior of the XC kernels computed for $r_s=2$.
Therefore, the considered case of $r_s=4$ more clearly illustrates the importance of the averaging procedure to reveal the correct trends. 

To more clearly analyse the  difference between the data for the individual snapshots and the  averaged values corresponding to these snapshots, we show $\Delta \chi (q)$ defined in Eq. (\ref{eq:d_chi_chi}), $\Delta \chi_{\rm KS} (q)$ defined in Eq. (\ref{eq:d_chiks_chiks}), and $\Delta K_{\rm xc} (q)$ computed using Eq. (\ref{eq:d_kxc_kxc}) in Fig. \ref{fig:diff_chi_rs4}. Additionally, we provide the standard deviations estimated using $N_s=10$ snapshots of $N=14$ particles (see grey areas).

From Fig. \ref{fig:diff_chi_rs4}, we see that the averaged values of the static density response function, the KS response function, and the static XC kernel obtained using snapshots with $N=14$ particles  exhibit a disagreement with the results from one snapshot with $N=112$ particles of less than $2\%$ (depicted using red squares).
In contrast, this difference (blue circles) reaches about $8\%$ for the density response function, about $17\%$ for the KS response function, and about $14\%$ for the static XC kernel for the considered individual snapshots with $N=14$ particles.

\subsection{Dynamic density response function}\label{s:results_B}

In order to demonstrate the application of Eq. (\ref{eq:chi_KS}) derived in Sec. \ref{s:theory_B} for averaging the dynamic KS response function, we present in Fig. \ref{fig:eps2} the results for the dynamic KS response functions $\chi_{\rm KS}(q,\omega)$ and $\chi_{{\rm KS},G=0}^{i}(q,\omega)$, and dynamic dielectric functions $\varepsilon_{\rm KS}(q,\omega)$ and $\varepsilon_{{\rm KS},G=0}^{i}(q,\omega)$ for $q/q_F\simeq 0.084$. 

In the top rows of Fig. \ref{fig:eps2}, we show the real and imaginary parts of $\chi_{{\rm KS},G=0}^{i}(q,\omega)$ (grey lines).
For the calculation of $\chi_{{\rm KS},G=0}^{i}(q,\omega)$ we used Eq.~(\ref{eq:macro_ks}). 
We also show $\chi_{\rm KS}(q,\omega)$ (blue lines) calculated using data from $N_s=250$ snapshots of $N=14$ particles according to  Eq. (\ref{eq:chi_KS_v2}). 
For the calculation of $\chi_{G=0}^{i}(q,\omega)$, we used Eq. (\ref{eq:chi_i}), with the static XC kernel $K_{\rm xc}^{i}(q)$ being defined by Eq. (\ref{eq:invert_i}). 

The second row from the top in Fig.\ref{fig:eps2}  shows the normalized difference between the $\chi_{\rm KS}(q,\omega)$ and the  mean arithmetic value $ \braket{\chi_{\rm KS}(\mathbf{q},\omega)}= \frac{1}{N_s}\sum_{i=1}^{N_s}\chi_{\rm KS,\vec G=0}^{i}(\mathbf{q},\omega)$. The normalized difference between the real and imaginary parts of $\chi_{\rm KS}(q,\omega)$ and $\braket{\chi_{\rm KS}(\mathbf{q},\omega)}$ are computed as
\begin{equation}
    \Delta {\rm Re}~{\chi}_{\rm KS}=\frac{{\rm Re}\braket{\chi_{\rm KS}(\mathbf{q},\omega)}-{\rm Re}~\chi_{\rm KS}(q,\omega)}{{\rm max}\left|{\rm Re} ~\chi_{\rm KS}(q,\omega)\right|}\times 100\%,
\end{equation}
and 
\begin{equation}
    \Delta {\rm Im}~{\chi}_{\rm KS}=\frac{{\rm Im}\braket{\chi_{\rm KS}(\mathbf{q},\omega)}-{\rm Im}~\chi_{\rm KS}(q,\omega)}{{\rm max}\left|{\rm Im}~ \chi_{\rm KS}(q,\omega)\right|}\times 100\%.
\end{equation}

The third row from the top corresponds to the real and imaginary parts of the dynamic dielectric functions $\varepsilon(q,\omega)$ (blue lines) and $\varepsilon^{i}(q,\omega)$ (grey lines); with $\varepsilon^{i}(q,\omega)$ being defined by Eq. (\ref{eq:eps_i}). To find  $\varepsilon(q,\omega)$ according to Eq. (\ref{eq:eps}), we  calculated the averaged value $\chi(q,\omega)$  by combining $\chi_{\rm KS}(q,\omega)$ with $K_{\rm xc}(q)$ (defined by Eq. (\ref{eq:invert}))  using Eq. (\ref{eq:chi_adiabatic}). 

Finally, the bottom rows shows the the normalized difference between the real and imaginary parts of the $\varepsilon(q,\omega)$ and the  mean arithmetic value $ \braket{\varepsilon(q,\omega)}= \frac{1}{N_s}\sum_{i=1}^{N_s}\varepsilon^{i}(q,\omega)$  computed as
\begin{equation}
    \Delta {\rm Re}~\varepsilon=\frac{{\rm Re}\braket{\varepsilon(q,\omega)}-{\rm Re}~\varepsilon(q,\omega)}{{\rm max}\left|{\rm Re}~ \varepsilon(q,\omega)\right|}\times 100\%,
\end{equation}
and 
\begin{equation}
    \Delta {\rm Im}~\varepsilon=\frac{{\rm Im}\braket{\varepsilon(q,\omega)}-{\rm Im}~\varepsilon(q,\omega)}{{\rm max}\left|{\rm Im}~ \varepsilon(q,\omega)\right|}\times 100\%.
\end{equation}

From  Fig. \ref{fig:eps2}, we see that  the $\Delta {\rm Re}~{\chi}_{\rm KS}$ and $\Delta {\rm Re}~\varepsilon$ values reach up to about $1\%$  at $q/q_F\simeq 0.084$. For the $ \Delta {\rm Im}~{\chi}_{\rm KS}$ and $\Delta {\rm Im}~\varepsilon$, we found the largest deviation values about $2\%$.
We note that the  standard error ( a standard deviation divided by $\sqrt{N_s}$) is smaller than the observed values of the difference of the results obtained using different averaging formulas. This dependence on the averaging formulas diminishes with the increase in the wavenumber as it is illustrated in Appendix \ref{sec:app_C} for $q/q_F\simeq 0.758$.
This makes intuitive sense as smaller values of $q$ correspond to the probing of larger length, leading to an increase in finite-size effects without proper averaging. 

In general, it is clear that the values of the errors due to  an inconsistent averaging over snapshots depend on the characteristics of the system under consideration and can be both smaller or larger than that of in the considered example of partially degenerate warm dense hydrogen. The usage of the presented  averaging workflow allows one to eliminate this unnecessary uncertainty.


\section{Conclusions and Outlook}\label{s:end}

We have presented a consistent scheme for the computation of the properly averaged macroscopic dynamic dielectric function and KS response function.
We used the adiabatic (static) approximation to $\chi(q, \omega)$, which is based on the static XC kernel calculation method developed recently in Refs. \cite{moldabekov_lr-tddft_2023, Moldabekov_dft_kernel} using the direct perturbation approach. The strength of this method is that it allows one to compute the static XC kernel for any available XC functional. 

For disordered systems, the  dynamic density response function and dynamic dielectric function depend on the positions of ions in the used snapshot. This dependence, together with periodic boundary conditions,  represents a finite size effect since in the extended macroscopic disordered systems a structure of ions (atoms) does not have periodicity as in crystals. 
Since the size of the main simulation cell is proportional to the inverse cube root of the number of particles in it,
the increase in the number of particles is not an effective strategy for computationally expensive  \textit{ab intio} simulation methods like thermal KS-DFT (particularly at high temperatures \cite{Fiedler_PRR_2022}), the generalized KS-DFT employing hybrid XC functionals \cite{Seidl_PRB, RevModPhys.80.3, PhysRevX.10.021040, Moldabekov_non_empirical_hybrid, hybrid_results}, and quantum Monte Carlo methods \cite{review}. Alternatively, one can perform averaging over snapshots to diminish this finite size effect. 
We have demonstrated that this can be an effective strategy on the example of warm dense hydrogen using $N_s=10$ snapshots of $N=14$ particles.   

Furthermore, considering an external perturbation with a wavenumber $\vec q$, we have shown that the induced density and KS potential excitations at $\vec q +\vec G$ (where $\vec G\neq 0$) disappear after averaging over snapshots. Therefore, we have demonstrated that a sufficiently weak  external harmonic perturbation  induces the linear density response only at the same wavenumber $\vec q$ as that of the external harmonic field for disordered systems. 
If one increases the amplitude of the perturbing field, the response of the system becomes non-linear \cite{Dornheim_PRR_2021, Moldabekov_JCTC_2022,Dornheim_PRL_2020}. This is known to be manifested by the appearance of the density excitations at higher harmonics $2q$ (for the quadratic response), $3q$ (for the cubic response) etc.  In a recent paper by B\"ohme \textit{et al} \cite{Bohme_PRL_2022}, it was shown for warm dense hydrogen  using quantum Monte Carlo simulations  (at parameters similar to those considered in this work), that it is problematic to resolve  a non-linear excitations at different harmonics for a single snapshot with $14$ or $24$ particles. 
In the present work, we have demonstrated the generation of nonphysical density excitations at $\vec G\neq0 $ in warm dense hydrogen  due to finite size effects.  These excitations at $\vec G\neq0 $  can overlap with the contributions from the non-linear density responses  generated at higher harmonics. 
Indeed, we have shown that such nonphysical density excitations at $\vec G\neq0 $ diminish after  averaging over snapshots .  Therefore, the averaging is essential for the simulation of the non-linear response properties of disordered systems.

 We have demonstrated that the calculation of the static XC kernel using a proper averaging procedure is important for obtaining adequate data for the XC kernel, and for the analysis of its properties. We stress that, in addition to its application in LR-TDDFT, the static XC kernel is important for a great variety of other applications such as the computation of effective  interaction potentials between particles \cite{Dornheim_Force_2022, zhandos_cpp21, zhandos_cpp17, zhandos2}, energy loss characteristics of dense plasmas \cite{Moldabekov_PRE_2020}, and for the application within time-dependent orbital-free DFT \cite{doi:10.1063/5.0100797} and  quantum hydrodynamics \cite{zhandos_pop18, Moldabekov_SciPost_2022, Graziani_CPP_2022}.   

Finally, we note that the inverse value of the  macroscopic static KS response function is connected to the second order functional derivative of the non-interacting free energy functional (kinetic energy functional at $T=0$) via the stiffness theorem \cite{quantum_theory, zhandos_pop18}. This relation is used to construct non-interacting free energy functionals for orbital-free DFT (OF-DFT) applications for both condensed matter \cite{wang2002orbital, moldabekov_ofdft_2023} and warm dense matter applications \cite{Sjostrom_PRB2013, PhysRevB.98.144302, PhysRevLett.121.145001}. In prior works, the UEG limit at which the  macroscopic KS response function reduces to the Lindhard function was used for the construction of non-interacting free energy functionals \cite{MGP, LMGP, PGSL, Sjostrom_PRB2013}. 
The recipe presented in this work for the computation of the macroscopic static KS response function  of real materials using the direct perturbation approach allows one to design more advanced non-interacting free energy functionals. In this way, the presented method for the macroscopic static KS response function is of  relevance for other DFT applications beyond WDM.

\section*{Acknowledgments}
This work was funded by the Center for Advanced Systems Understanding (CASUS) which is financed by Germany’s Federal Ministry of Education and Research (BMBF) and by the Saxon state government out of the State budget approved by the Saxon State Parliament. This work has received funding from the European Research Council (ERC) under the European Union’s Horizon 2022 research and innovation programme
(Grant agreement No. 101076233, "PREXTREME"). We gratefully acknowledge computation time at the Norddeutscher Verbund f\"ur Hoch- und H\"ochstleistungsrechnen (HLRN) under grant shp00026, and on the Bull Cluster at the Center for Information Services and High Performance Computing (ZIH) at Technische Universit\"at Dresden.

\appendix

\section*{Appendix A\label{sec:app_A}}
Within linear response theory for homogeneous systems, the density response function is expressed in terms of the non-interacting density response function $\chi_0$ and the exchange-correlation kernel as \cite{quantum_theory, book_Ullrich}
\begin{eqnarray}\label{eq:chi_app_A}
 \chi(\mathbf{q},\omega) = \frac{\chi_{0}(\mathbf{q},\omega)}{1 -\left[v(q)+K_{\rm xc}(\mathbf{q})\right]\chi_{0}(\mathbf{q},\omega)}.
\end{eqnarray}

In the KS-DFT framework, the non-interacting density response function is given by the KS response function, i.e., 
$\chi_{0}=\chi_{\rm KS}$. For the static case, if one substitutes Eq. (\ref{eq:chi_KS}) for the KS response function into  Eq. (\ref{eq:chi_app_A}) (with $\omega=0$), Eq. (\ref{eq:my_chi}) is reproduced for the averaged value of the density response function. 

In contrast, the alternative definition of the  KS response function Eq. (\ref{eq:i_chi_KS}) is not compatible with  Eq. (\ref{eq:chi_app_A}). Let us for simplicity consider the case of RPA, i.e, $K_{\rm xc}=0$. We first express the KS response function for a given snapshot in terms of the density response function by inverting Eq. (\ref{eq:chi_i}) with $K_{\rm xc}=0$:

\begin{equation}\label{eq:chi_i_invert}
    \chi_{\rm KS, \vec G=0}^{i}(\mathbf{q})=\frac{\chi_{\vec G=0}^{i}(\mathbf{q})}{1+v(q)\chi_{\vec G=0}^{i}(\mathbf{q})}\ ,
\end{equation}
where $\chi_{\rm KS, \vec G=0}^{i}(\mathbf{q})=\chi_{\rm KS, \vec G=0}^{i}(\mathbf{q},\omega=0)$ and $\chi_{\vec G=0}^{i}(\mathbf{q})=\chi_{\vec G=0}^{i}(\mathbf{q},\omega=0)$.
Using Eq. (\ref{eq:chi_i_invert}) in Eq. (\ref{eq:i_chi_KS}) , we get:
\begin{equation}\label{eq:chi_A_m}
    \braket{\chi_{\rm KS}(\mathbf{q})}= \frac{1}{N_s}\sum_{i=1}^{N_s}\frac{\chi_{\vec G=0}^{i}(\mathbf{q})}{1+v(q)\chi_{\vec G=0}^{i}(\mathbf{q})}
\end{equation}
Inserting Eq. (\ref{eq:chi_A_m}) and Eq. (\ref{eq:my_chi}) into Eq. (\ref{eq:chi_app_A}) then leads to the inequality
\begin{equation}
    \chi(\mathbf{q}) =\frac{1}{N_s}\sum_{i=1}^{N_s} \chi_{\vec G=0}^{i}(\mathbf{q})\neq \frac{\braket{\chi_{\rm KS}(\mathbf{q})}}{1 -v(q)\braket{\chi_{\rm KS}(\mathbf{q})}}.
\end{equation}

\section*{Appendix B\label{sec:app_B}}

\begin{figure}[htp!]\centering
\includegraphics[width=0.45\textwidth]{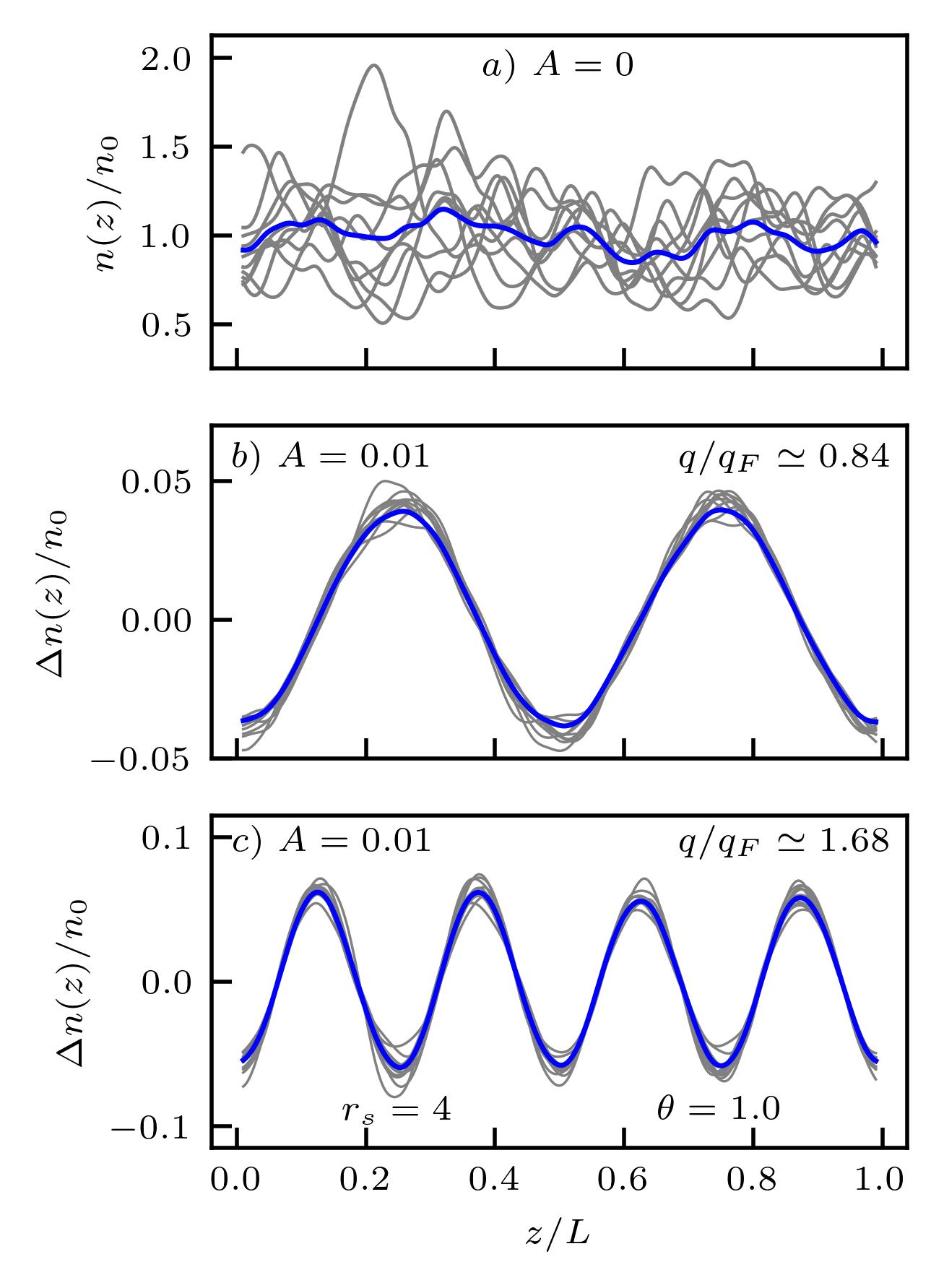}
\caption{\label{fig:den_rs4_N112} a) Density distribution along the z axis of the unperturbed system, b) density perturbation at $A=0.01$ and $q\simeq 0.84q_F$, and  c) density perturbation at $A=0.01$ and $q\simeq 1.68q_F$. The results are computed for $N=112$ particles with $r_s=4$ and $\theta=1$. 
}
\end{figure}

In Fig. \ref{fig:den_rs4_N112}, we show the density profiles for $N_s=10$ snapshots with $N=112$ particles (solid grey lines) at $r_s=4$ and $\theta=1$ in the cases of the unperturbed and perturbed dense hydrogen gas. From Fig. \ref{fig:den_rs4_N112} a), we observe that the density deviations from the mean value are effectively reduced due to averaging over snapshots. 
From Fig. \ref{fig:den_rs4_N112}b) and Fig. \ref{fig:den_rs4_N112}c),  one can see that the averaging of the the density perturbation over 10 snapshots effectively reduces the deviations from the cosinuoidal profile.  

\begin{figure*}
\includegraphics[width=\linewidth]{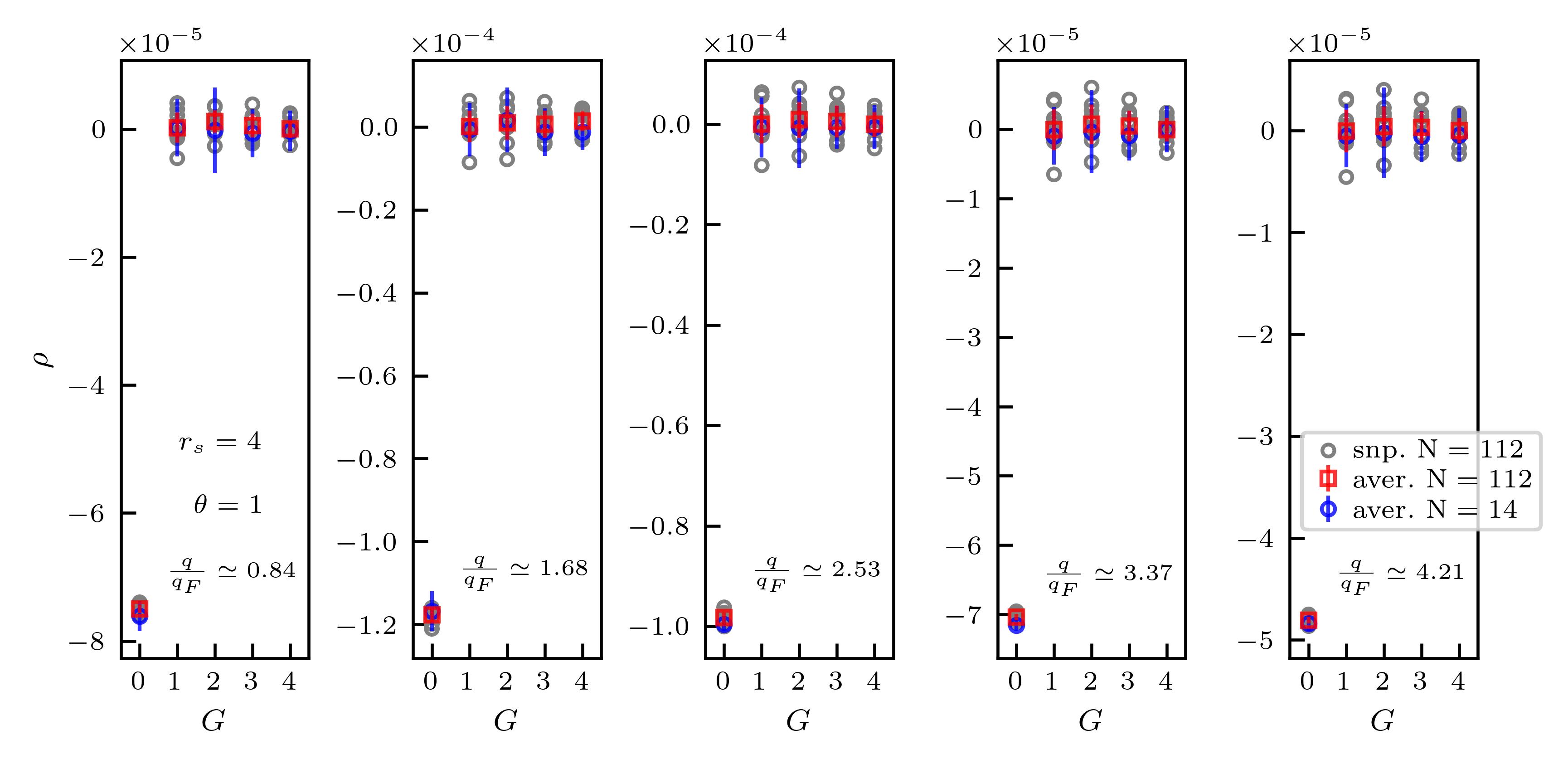}
\caption{\label{fig:rho_rs4_N112}  Contributions to the total density change from the density perturbation values at different wavenumbers for different snapshots (the grey circles are for 112 particles), and for the averaged values over 10 snapshots with 112 particles (red) in warm dense hydrogen at $r_s=4$ and $\theta=1$. Additionally, we show the averaged values over 10 snapshots with 14 particles (blue)
The wave number $q$ corresponds to the wavenumber of the external perturbation.  The $\vec G$ is along the z-axis and in units of $2\pi/L$.
}
\end{figure*}

In Fig. \ref{fig:rho_rs4_N112}, we show the density perturbation values $\rho_{G}(q)$ (as defined in Eq. (\ref{eq:expansion})) at different $G$ and $q$ values for 112 particles. We see that the contribution of  $\rho_{G\neq0}(q)$ components cancel each other after averaging over snapshots. 

\section*{Appendix C\label{sec:app_C}}

\begin{figure*}[t!]
\includegraphics[width=\linewidth]{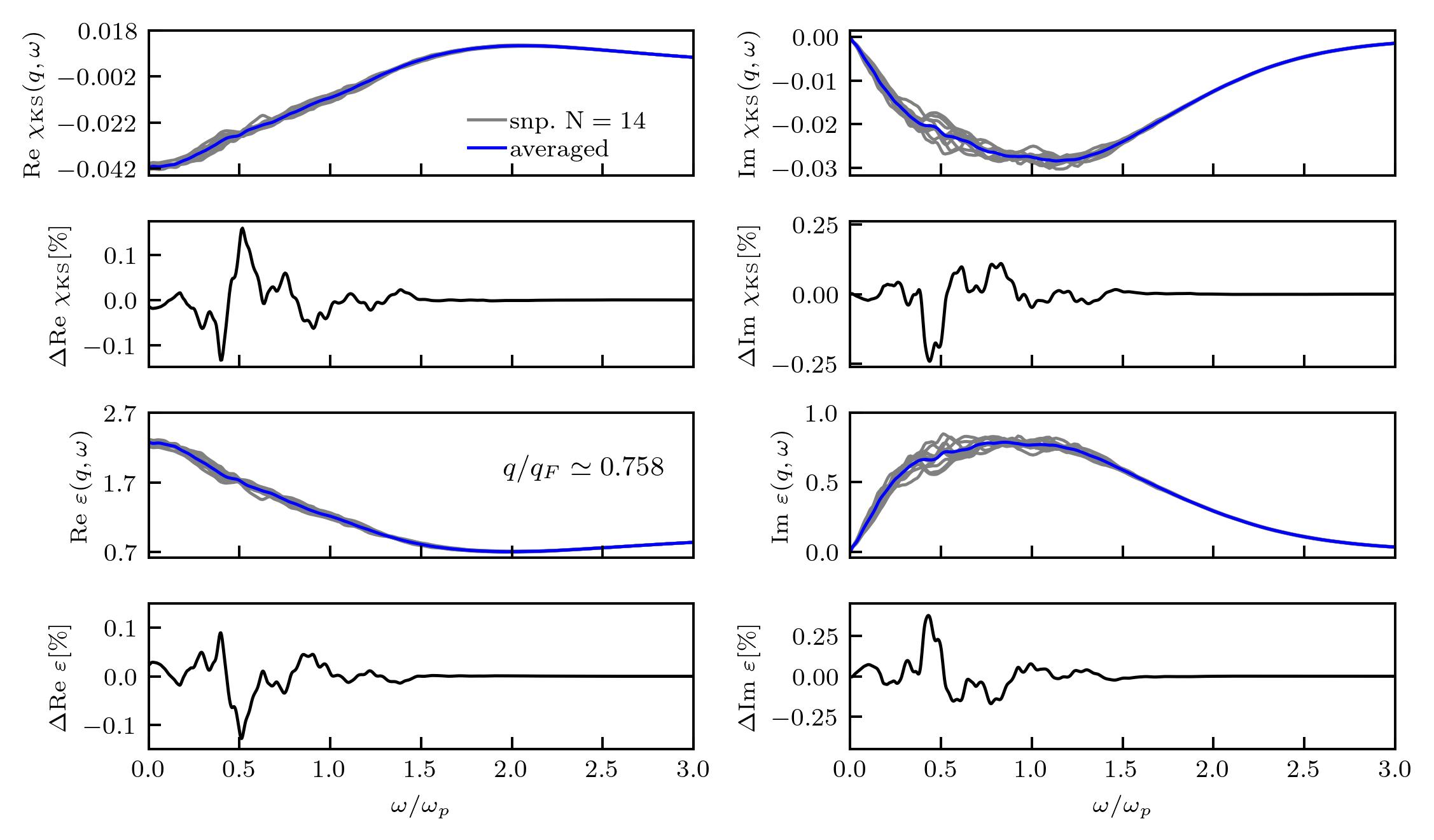}
\caption{\label{fig:eps1}   First row: real and imaginary part of the macroscopic KS response function. 
Second row:  difference between the results for the KS response function computed using different averaging methods. 
Third row: real and imaginary  part of the dielectric function.
Bottom row:  difference between the results for the macroscopic dielectric function computed using different averaging methods. 
The results for a given snapshot are presented by grey lines (with 14 particles). 
The averaged values over 10 snapshots  are presented by blue lines. The results  are computed for warm dense hydrogen at $r_s=2$ and $\theta=1$ with $q/q_F\simeq 0.758$. We note that the difference between the results  computed using different averaging methods is within standard error and negligible.
}
\end{figure*}

In Fig. \ref{fig:eps1} the results for the dynamic KS response functions $\chi_{\rm KS}(q,\omega)$ and $\chi_{{\rm KS},G=0}^{i}(q,\omega)$, and dynamic dielectric functions $\varepsilon_{\rm KS}(q,\omega)$ and $\varepsilon_{{\rm KS},G=0}^{i}(q,\omega)$ for $q/q_F\simeq 0.758$ are shown. 
The second and  bottom rows show the difference in the results computed using different averaging formulas as it is discussed in Sec. \ref{s:results_B}.
We see that the difference between different considered averaging formulas are negligible for $q/q_F\simeq 0.758$.

\bibliography{bibliography.bib}

\end{document}